\documentclass[urlcolor=cyan, colorlinks=true, citecolor=blue, linkcolor=blue]{aa}
\usepackage[varg]{txfonts}
\usepackage{placeins}
\usepackage{linenoaa}
\usepackage{lscape}
\usepackage{titlesec}
\usepackage{orcidlink}
\usepackage{hyperref}
\bibpunct{(}{)}{;}{a}{}{,}

\begin{document}

\title{The Cygnus Allscale Survey of Chemistry and Dynamical Environments: CASCADE }
\subtitle{VI. Molecular outflows in the DR21 ridge}
\author{I.~M.~Skretas\inst{1}, A. Karska\inst{2,1}\orcidlink{0000-0001-8913-925X}, F.~Wyrowski\inst{1}, H.~Beuther\inst{3}, W.-J.~Kim\inst{1}, C.~Gieser\inst{3}, D.~A.~Semenov\inst{4,3}, A.~Hernández-Gómez\inst{5}, N.~Schneider\inst{6}, T.~Csengeri\inst{7}, A.~Hacar\inst{8}}

\institute{Max-Planck-Institut für Radioastronomie, Auf dem Hügel 69, 53121, Bonn, Germany
\and Institute of Advanced Studies, Nicolaus Copernicus University in Toruń, Wileńska 4, 87-100 Toruń, Poland
\and Max Planck Insitute for Astronomy, Königstuhl 17, 69117 Heidelberg, Germany
\and Zentrum f\"{u}r Astronomie der Universit\"{a}t Heidelberg,
Institut f\"{u}r Theoretische Astrophysik, Albert-Ueberle-Str. 2,
69120 Heidelberg, Germany
\and Tecnologico de Monterrey, Escuela de Ingeniería y Ciencias, Avenida Eugenio Garza Sada 2501, Monterrey 64849, Mexico
\and I. Physik. Institut, University of Cologne, Cologne, Germany
\and Laboratoire d’astrophysique de Bordeaux, Univ. Bordeaux, CNRS, B18N, allée Geoffroy Saint-Hilaire, 33615 Pessac, France
\and Department of Astrophysics, University of Vienna, Türkenschanzstraße 17 (Sternwarte) 1180 Wien, Austria}

\date{Received 08 June 2026; Accepted }
\titlerunning{Outflows in Cygnus-X North}
\authorrunning{I.~Skretas et al. }

\abstract
{Star formation takes place in varied environments, from isolated clumps to massive molecular cloud complexes. However, whether the environment in which a star forms has any effect on the formation process remains a matter of debate.
The molecular outflows, launched during the formation of protostars can present a more easily accessible way to study star formation in different environments.
The DR21 ridge, located in the Cygnus-X high-mass star-forming complex, represents a well-known example of ongoing high-mass star formation, hosting a high number of massive dense cores and embedded protostars with outflows.}
{We aim to identify molecular outflows associated with dense molecular cores, the earlier stages of high-mass star formation, along the DR21 ridge, and investigate whether the extended environment of a dense core impacts the formation process of stars within it.}
{We identified molecular outflows along the entire length of the DR21 ridge using HCO$^+$ $J=1-0$, H$^{13}$CO$^+$ $J=1-0$, and SiO $J=2-1$ observations obtained with the Institut de Radioastronomie Millimètrique (IRAM) 30 m telescope and the Northern Extended Array for Millimeter Astronomy (NOEMA) as part of the Cygnus Allscale Survey of Chemistry and Dynamical Environments (CASCADE) program. We calculated outflow properties (i.e. the outflow mass, $M_\text{out}$, momentum, $p_\text{out}$, kinetic energy, $E_\text{kin}$, mass loss rate, $\dot{M}_\text{out}$, force $F_\text{out}$, and kinetic luminosity, $L_\text{kin}$) and performed statistical comparisons between the DR21 ridge sources and an extended literature sample of outflow sources, ranging from low- to high-mass sources. }
{Based on the morphology of HCO$^+$, H$^{13}$CO$^+$ and SiO, we identify molecular outflows in 14 out of 34 dense cores (41\,\%) along the DR21 ridge. Despite the high density of star formation in DR21, the resulting outflow properties are found to be in good agreement with the established correlations between outflow and source properties (i.e. envelope mass, $M_\text{env}$, and bolometric luminosity, $L_\text{bol}$). In addition, little variation is seen in the outflow properties of sources along the ridge, with the exception of the sources that are located at the intersection of the DR21 ridge with large-scale ($\sim$1 pc) accretion filaments. These sources are found to drive the most powerful outflows.}
{Overall, our results indicate that protostellar outflow properties, even when driven by sources forming in an extreme and clustered region, such as the DR21 ridge, remain largely unaffected. In turn, the close connection between outflows and the accretion process suggests that star formation as a whole remains mostly unchanged, despite large variations of the large-scale environment in which it takes place, and is rather governed by more local processes.}

\maketitle
\nolinenumbers

\section{Introduction}
Bipolar outflows serve as a clear signpost of star formation. They are a necessary by-product of the accretion process, as by removing excess angular momentum from the protostar-disk system they enable the infall of material along the disk \citep{Pudritz07,Frank2014}. Star formation models describing the launch of protostellar outflows initially proposed for low-mass stars \citep[e.g.,][]{Shu94,Pudritz1983} have also been developed for the high-mass counterparts \citep[e.g.,][]{Vaidya2011,Oliva2023}, and such outflows have been detected in sources across the mass spectrum \citep[e.g.,][]{Beuther2005,Arce07,bally2016}. These outflows inject significant amounts of momentum and energy in the surrounding interstellar medium (ISM) and are believed to be responsible for the dispersion of the natal, protostellar envelopes. As a result, protostellar outflows play a key role in regulating the star formation efficiency \citep[e.g.,][]{Fall10,Frank2014}. 

Protostellar outflows are most often observed in the low $J$ rotational transitions of CO \citep[e.g.,][]{vdm13}, but other molecules, such as HCO$^+$ \citep[e.g.,][]{Girart1999,Arce2006,Tafalla2010,WS14,Podio2014,Skretas23} and SiO \citep[e.g.,][]{Mot07,duarte14,Yang24} have also been used albeit with higher uncertainties due to significant variations observed in their abundance ratios \citep[e.g.,][]{Godard2010,Gerner2014}. These molecules are tracing the colder, entrained molecular gas and allow for the calculation of several energetic properties of the protostellar outflow, such as the outflow mass $M_\text{out}$, momentum $p_\text{out}$ and kinetic energy $E_\text{kin}$, as well as the equivalent rates, mass loss rate $\dot{M}_\text{out}$, force $F_\text{out}$ and kinetic luminosity $L_\text{kin}$. Correlations, connecting these outflow properties with the properties of the molecular core (e.g., the mass and luminosity) hosting the driving protostar, have been discovered, both for low- \citep[e.g.,][]{Bontemps1996,Yildiz2015,Mottram2017} and high-mass sources \citep[e.g.,][]{Beuther2002,Maud2015}. 
Recent studies of these correlations across extended mass regimes have shown significant differences between the behavior of low- and high-mass sources, with low-mass sources displaying much higher scatter compared to their high-mass counterparts \citep{Skretas22}. These differences are further highlighted by the limited number of intermediate-mass sources \citep[e.g.][]{vk16} included in these studies. The overall similarities and differences of low- and high-mass star formation remain a matter of discussion \citep[e.g.,][]{Beuther2025}.  
In addition, previous studies, especially in the high-mass regime, often targeted sources in different molecular clouds with varying environments. By constraining the environment of the sources, the characteristics of the driving sources that influence the outflow properties can better be traced.

In this work, we aim to identify and characterize all molecular outflows associated with dense molecular cores, the early stage of high-mass star formation, within the DR21 ridge in the Cygnus X region. Using these sources, we intend to expand the established correlations between the properties of outflows and their driving sources and investigate whether these are impacted by the environment in which the stars are forming.  More precisely, the goal is to compare the outflow properties of sources along the DR21 ridge with the established correlations in order to identify whether the extreme nature of the filament impacts the formation process of stars.

To that end, we present observations of the DR21 ridge, taken as part of the Cygnus Allscale Survey of Chemistry and Dynamical Environments (CASCADE) \citep{Beuther2022}. CASCADE is a large Max Planck IRAM Observatory Program (MIOP) covering the densest regions of the Cygnus-X molecular cloud complex in the 3\,mm window, with high angular resolution ($\sim 3\arcsec$ corresponding to $\sim$ 4000 AU) and with a broad bandpass. The observations were carried out using both the Northern Extended Array for Millimeter Astronomy (NOEMA) and the 30 m telescope, operated by the Institut de Radioastronomie Millimètrique (IRAM). The combined use of single-dish and interferometric observations is crucial as it provides the angular resolution required to identify molecular outflows while at the same time, maintaining the full information of the more extended scales. For more details on the scope and goals of CASCADE we refer the reader to the overview publication of the project by \cite{Beuther2022}. 

The Cygnus-X giant molecular cloud complex, located at a distance of $\sim1.4$ kpc \citep{Rygl2012} is one of the nearest active high-mass star-forming regions in the Milky Way. It is also one of the largest molecular cloud complexes, with an estimated total molecular mass of $\sim3 \times 10^6$ M$_\odot$ and a projected size of $\sim130$ pc \citep{Schneider2006,Reipurth2008,Cao19}. The complex hosts all stages of high-mass star-formation, containing a large number of ultra-compact H{\sc ii} regions \citep{Cyg03} and a significant population of OB stars \citep{wright2010}, along with one of the largest known OB associations in our Galaxy, Cyg OB2 \citep{knodlseder2000}.   

\begin{table}[htb!]
\caption{Continuum and spectral line parameters in this work} 
\label{table:observations} 
\centering 
\small
\setlength{\tabcolsep}{4pt}
\footnotesize{
\begin{tabular}{l c c c c c c} 
\hline\hline 
Tracer & Trans. & \begin{tabular}[c]{@{}c@{}}Freq.\\  {[}GHz{]}\end{tabular}  & \begin{tabular}[c]{@{}c@{}}$\sigma_\text{rms}$\\  $\left[\frac{\rm \text{mJy}}{\rm \text{beam}}\right]$ \end{tabular} & \begin{tabular}[c]{@{}c@{}}Beam\\  {[}arcsec{]}\end{tabular} & \begin{tabular}[c]{@{}c@{}}$E_\mathrm{up}/k$\\  {[}K{]}\end{tabular} &  \begin{tabular}[c]{@{}c@{}}log$_\text{10}$($A_\mathrm{ij}$)\\  {[}s$^{-1}${]}\end{tabular} \\ 
\hline
3~mm cont. & -- & 90.0& 0.07& 3.32$\times 2.66$ & -- & --\\
H$^{13}$CO$^+$ & 1-0 & 86.754& 7.5 & 2.74$\times$2.49 & 4.16& -4.41 \\
SiO & 2-1 & 86.847& 9.6 & 3.25$\times$2.70 & 6.25& -4.53  \\
HCO$^+$ & 1-0 & 89.189& 7.7 &3.29$\times$2.74 & 4.28& -4.38  \\
\hline 
\end{tabular}}
\tablefoot{
The rms noise level, $\sigma_\text{rms}$, is calculated for channels of 2.0 km s$^{-1}$. Rest frequencies, upper level energies, E$_\mathrm{up}$, and the Einstein coefficients for spontaneous emission, A$_\mathrm{ij}$ are taken from the Cologne Database for Molecular Spectroscopy (CDMS, \citet{Muller2001}). The conversion factor from Jy/beam to K, in the Rayleigh-Jeans limit, is $\sim$ 23.75 for a frequency of 89.189 GHz.}
\end{table} 

The DR21 ridge is the most active region within Cygnus-X thus offering an ideal target to study the impact of a clustered environment on the process of star formation. It spans a length of $\sim 8$ pc with a total mass of $\sim 10^4$ M$_\odot$ and typical H$_2$ column densities of $\sim 10^{23}$ cm$^{-2}$ \citep{Schneider10,Hen12}. At the southern end of the DR21 ridge is the prominent compact H~{\small II} region DR21 Main \citep{DownesR66,Immer14} driving a massive outflow perpendicular to the filament \citep{Gar86,Gar91,Skretas23,Karska2025} believed to be, at least in part, of explosive rather than protostellar origin \citep{Zapata2013,Guzman2024}. The well studied OH, H$_2$O, and CH$_3$OH maser source DR21 (OH) is located further north \citep{mangum1992,Karska2014, Orozco2019}. In addition, \cite{Mot07} using IRAM 30m telescope observations at 1.2 mm identified a significant number of dense molecular cores along the DR21 ridge. These cores display a gradient of evolutionary stages, with the northern end of the ridge being less evolved than the southern end \citep{Hen12}. The list of dense cores in DR21 was expanded by \cite{Cao19,Cao21}, who made use of additional continuum observations of the complex at shorter wavelengths (70 - 850 $\mu$m) taken with $Herschel$ PACS and SPIRE instruments, as well as, with the James Clerk Maxwell Telescope (JCMT) SCUBA receiver. In connection to these cores, several molecular outflows have been identified along the ridge, primarily through CO and SiO observations \citep[e.g.,][]{Mot07,Schneider10,duarte14,Skretas22,Yang24}, while H$^{13}$CO$^{+}$ and H$^{13}$CN have been used in some cases to analyze the kinematics of denser gas \citep[e.g.,][]{Csengeri2011a}.        

The paper is organized as follows. Section~\ref{sec:observations} describes the observations from CASCADE as well as the different dense core catalogs covering the DR21 ridge. Section~\ref{sec:results} presents HCO$^+$ and H$^{13}$CO$^+$ $J=1-0$ spectra along with the associated integrated intensity maps for the high-velocity HCO$^+$ emission for all cores along DR21. In addition, outflow properties for all detected outflows are calculated. In Section~\ref{sec:discussion} the derived outflow properties are discussed both in context with the known properties of the DR21 ridge but also in comparison with additional literature samples. A brief comparison between HCO$^+$ and CO outflows from previous studies is also presented in this section. Finally, Section~\ref{sec:conclusions} contains the summary and conclusions.
 
\section{Observations and data analysis}
\label{sec:observations}
The observations of the DR21 ridge, presented in this work, were taken as part of the CASCADE project \citep{Beuther2022}. 
The DR21 ridge was covered by a total of six NOEMA mosaic tiles, each covering an area of 16 arcmin$^2$ and corresponding to 78 NOEMA pointings. The separation between pointings is set to 27 arcseconds, corresponding to Nyquist sampling at a frequency of 90GHz. Observations were carried out using both the C and D configurations and have a total bandwidth of 16 GHz in the 3~mm atmospheric window. The spectral resolution in the full frequency range probing the continuum is 2.0 MHz (or $\sim6.5$ km s$^{-1}$), while prominent spectral lines are covered in dedicated narrow windows with a spectral resolution  of 62.5 kHz (or $\sim0.2$ km s$^{-1}$), thanks to the use of the additional high-resolution correlator units available to NOEMA. The NOEMA observations for the DR21 ridge were carried out between May 17 and November 18, 2020, with a total of 10 antennas available at the time, leading to baselines between $\sim$15 m and $\sim$ 365 m. For the bandpass calibration, the quasars 3C345 and 3C273 were used, for the flux calibration MWC349 and 2010+723, while for gain calibration 2005+403, 2120+445, 2050+363 and 2013+370 were used. We expect the total flux uncertainty to be approximately 10\%, due to the uncertainty of the total flux calibrators. For the calibration and imaging of the data, the CLIC and MAPPING software of the GILDAS package\footnote{\url{https://www.iram.fr/IRAMFR/GILDAS/}} were used.
 
Complementary single-dish observations, providing the missing short spacing information, were carried out with the IRAM 30 m telescope between February and July 2020. These observations are presented in detail in \cite{Ivalu2024}. The combination of NOEMA and 30m data was done during the imaging process using the task \texttt{UV\_SHORT}. To better cover the high-velocity material of the protostellar outflows, we reduced the spectral resolution of the combined NOEMA and IRAM 30m telescope data from the best possible resolution of 0.8 km s$^{-1}$ to 2 km s$^{-1}$, allowing us to create datacubes that cover extended velocity ranges with higher sensitivity. For this work, the HCO$^+$ $J = 1-0$, H$^{13}$CO$^+$ $J=1-0$ and SiO $J=2-1$ transitions are analyzed. The resulting single channel $\sigma_\text{rms}$ noise, for a channel width of 2 km s$^{-1}$, and beam sizes for the three targeted lines are shown in Table \ref{table:observations}. 

\subsection{Continuum observations}

\begin{figure}
    \centering
    \includegraphics[width=0.8\linewidth]{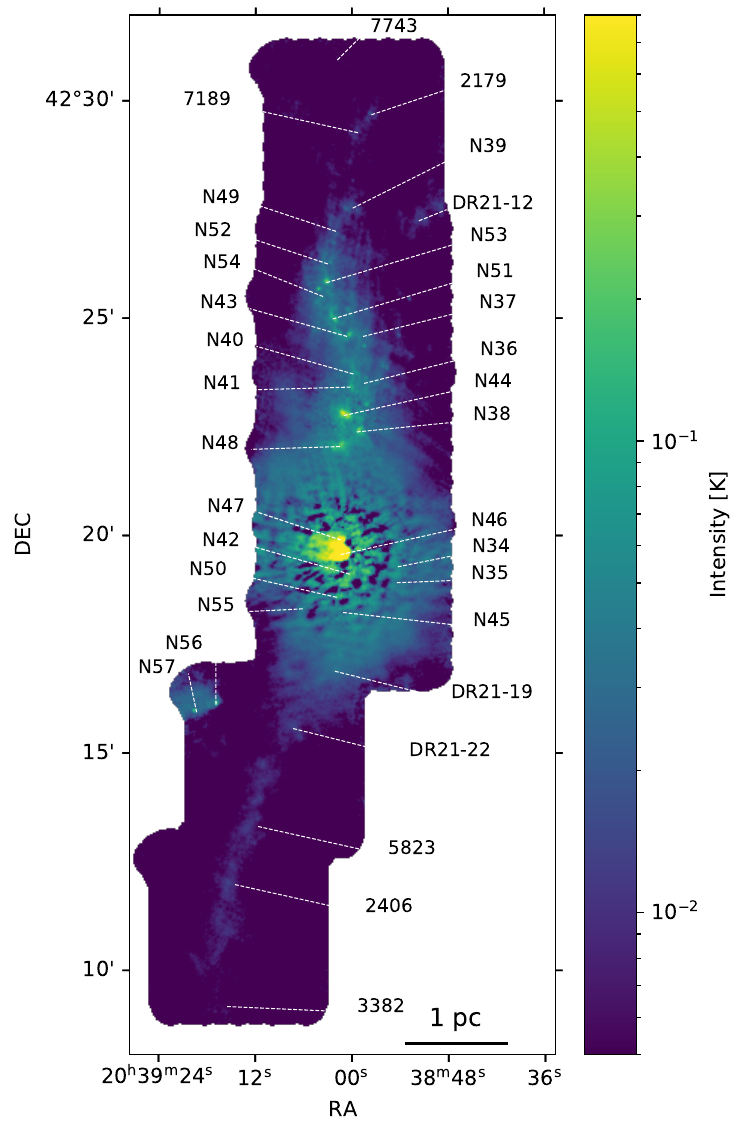}
    \caption{Combined NOEMA and GBT 3.6 mm continuum map of the DR21 ridge region. Marked are the location of all dense clumps associated with the region (see Sec. \ref{sec:densecores}). Sources named as N followed by a number (e.g., \lq N51') are taken from \cite{Mot07}, sources named as DR21 followed by a number (e.g., DR21-12) are taken from \cite{Cao19}, while cores named only with a four digit number (e.g., \lq 5823') are from \cite{Cao21}.}
    \label{fig:DR21_cont}
\end{figure}

From the full frequency range of the NOEMA observations, omitting spectral lines, the 3 mm continuum emission was constructed. In order to provide short-spacings information for the continuum, additional observations using the 100m Green Bank Telescope's (GBT) MUSTANG-2 instrument were performed. MUSTANG-2 offers a 4.2$'$ field-of-view (FOV), using an array of 215 feedhorn-coupled TES bolometers \citep{Dicker2014}, and has a 30 GHz bandpass, with an effective center close to 90 GHz \citep{Ginsburg2020}.    
The GBT observations were carried out between February 2 and March 14 2022, as part of the GBT22A-280 project (PI: A. Ginsburg). Data calibration and reduction was done using the MUSTANG IDL Data Analysis System \citep[MIDAS;][]{Romero2020}. 
The MUSTANG-2 observations were then combined with the upper side band data from NOEMA, which are also centered around 90 GHz. For the combination of the two datasets, the uv\_short task from the GILDAS software package was used. This uv\_short task creates a visibility dataset from the single-dish data that is then merged with the NOEMA data in the uv-plane. The merged dataset is then imaged classically in GILDAS using natural weighting. The final synthesized beam of the combined data is 3.32$\arcsec \times 2.66\arcsec$, and the noise is $\sigma_\text{rms} = 0.07$ mJy beam$^{-1}$ ($\sim 1.4$ K). The GBT observations along with the combination process are described in more detail in \cite{Kim2025}. The final continuum image for the entire DR21 ridge is shown in Fig. \ref{fig:DR21_cont}. Note that the strong artifacts, caused by the presence of the powerful continuum source DR21 Main (or N46), influence a large part of the mapped area. Since emission at 3.6~mm can also have strong free-free contributions these observations are not ideal for the identification of cores. Still, thanks to their higher angular resolution, they can assist in identifying the location of the driving sources of protostellar outflows within the known cores.   

\subsection{Dense molecular clumps in the CASCADE field}
\label{sec:densecores}

\begin{table*}
\caption{Core properties for sources in the DR21 CASCADE field} 
\label{table:cores} 
\centering 
\begingroup
\small 
\makebox[\textwidth][c]{
\begin{tabular}{l l c c c c c c c c c}
\hline \hline 
\# & Name & RA\tablefootmark{a} & DEC\tablefootmark{a}& \begin{tabular}[c]{@{}c@{}}$FWHM$\tablefootmark{a}\\  {[}pc{]}\end{tabular}& \begin{tabular}[c]{@{}c@{}}$M$\tablefootmark{a}\\  {[}M$_\odot${]}\end{tabular} & \begin{tabular}[c]{@{}c@{}}<n$_\text{H$_2$}$>\tablefootmark{a}\\  {[}10$^5$ cm$^{-3}${]}\end{tabular} & \begin{tabular}[c]{@{}c@{}}$T$\tablefootmark{b}\\{[}K{]}\end{tabular} & \begin{tabular}[c]{@{}c@{}}$N_\text{H$_2$}$\tablefootmark{c}\\  {[}10$^{22}$cm$^{-2}${]}\end{tabular}& \begin{tabular} [c]{@{}c@{}}$L_\text{FIR}$\tablefootmark{c}\\  {[}L$_\odot${]}\end{tabular} & Classification\tablefootmark{c}\\  
\hline
1& N36 & 20:38:58.49 & 42:23:30.12 &0.19& 11 &1.2 & -- & -- & -- & -- \\
2& N38/DR21-13 & 20:38:59.40 & 42:22:23.16 &0.18 / 0.054&138/ 85.3 & 2.7 / 131&20.7 & 152.7& 643& IR-quiet\\
3& N39/DR21-15& 20:38:59.90& 42:27:32.04& 0.16 / 0.18& 10 / 51.1& 0.3 / 2.1& 14.2& 8.1&40.1 &starless \\
4& N41& 20:39:00.19 & 42:23:25.08 & 0.03 & 10 & 19 & -- & -- & -- & -- \\
5& N42/DR21-17 & 20:39:00.29 & 42:19:06.96 & 0.05/0.206 & 5/47.6 & 4.5/1.3 & 20.6& 5.8 & 348 & IR-quiet\\
6& N45/DR21-16 & 20:39:01.10 & 42:18:14.04 & 0.13/0.159 & 16/50.6&0.64/3 & 15.9 & 10.3 & 79.9& IR-quiet \\
7& N47&20:39:01.39 & 42:19:54.12&0.08 & 19& --&-- & --& --&-- \\
8& N48/DR21-3& 20:39:01.49& 42:22:03& 0.17/0.177& 197/261& 3.5/11.1&22.6 &42.8 & 3353& IR-bright\\
9& N49/DR21-14& 20:39:01.99& 42:27:00& 0.15/(0.086)&8/72.6& 1.2/$\geq$32.1& 12.6& $\geq$56.4& 27.6& IR-quiet\\
10& N50/DR21-6\tablefootmark{1}& 20:39:01.90& 42:18:34.92&0.24/(0.086) &12/121 &1.0/$\geq$53.6 & 15.4& $\geq$94.3& 153& IR-quiet\\
11& \hspace{14pt} /DR21-20\tablefootmark{1}& -- & -- & 0.119 & 37.9&5.4&14.6&13.8&35.9&IR-quiet \\
12& N52/DR21-21& 20:39:03.00& 42:26:15& 0.16/0.175&33/37.2& 0.49/1.6&16.2& 6.3 & 64.8& IR-quiet\\
13& N53/DR21-7& 20:39:03.19& 42:25:49.08& 0.14/0.053& 85/115& 2.2/186.2& 17.0& 213.1& 267& IR-quiet\\
14& N54& 20:39:03.60& 42:25:30& 0.18& 24& 0.6& --& --& --&-- \\
15& N37& 20:38:58.61& 42:24:34.92& 0.03& 4& 8.7& --& --&-- & --\\
16& N40/DR21-8& 20:38:59.81& 42:23:43.08& 0.33/0.159& 106/114& 1.1/6.7& 19.8& 23.2& 672& IR-quiet\\
17& N43/DR21-23& 20:39:00.60& 42:24:34.92&0.19/0.106 &66/35.1&1.1/7 & 34.7& 16.1& 5865& IR-bright\\
18& N51/DR21-4& 20:39:02.40& 42:24:59.04& 0.19/0.078& 125/153&2.0/77.8 &19.7&131.3 &853 & IR-bright\\
19& N56/DR21-9& 20:39:16.90& 42:16:06.96& 0.12/0.053& 32/94.9&1.4/152.6 & 14.6& 175.2&89.7 & IR-quiet\\
20& N46/DR21-1& 20:39:01.39& 42:19:33.96& 0.19/0.195& 949/1355&9.1/43.3 &20.6 & 183.5&10095 & IR-bright\\
21& N44/DR21-2& 20:39:01.01& 42:22:45.84& 0.14/0.134& 446/1048&10.0/103.7 &21.0 &301.7 & 8600& IR-quiet\\
22& N55/DR21-11& 20:39:06.10& 42:18:19.08& 0.15/(0.086)&11/89.9 &0.8/$\geq$39.7 &14 & $\geq$69.8& 66.3& starless\\
23& N57/DR21-10& 20:39:19.30& 42:15:56.16& 0.09/0.06& 11/94.1&1.5/105.1 &10.3 &136.3 & 10.6& IR-quiet\\
24& N34/DR21-5& 20:38:54.29& 42:19:17.04& 0.19/0.226& 23/125&0.61/2.6 &17.5& 12.6&351 & IR-quiet\\
25& N35& 20:38:54.41& 42:18:55.08& 0.13& 8&1.5 & --& --& --&-- \\
26& DR21-19& 20:39:02.12& 42:16:53& 0.169& 39.6& 1.9& 16.2& 7.1& 68.6& IR-quiet\\
27& DR21-12& 20:38:51.65& 42:27:14.50&(0.086) & 88.5& $\geq$39.1& 13.0& $\geq$68.7& 41.7& IR-quiet\\
28& DR21-22& 20:39:07.30& 42:15:33.90& 0.202& 36.5& 1& 17.8& 4.6& 111& IR-quiet\\
29& 2179&20:38:57.567& 42:29:40.78& 0.272& 27.95& --& 17.4& --& --& --\\
30& 2406&20:39:14.526& 42:11:58.52& 0.559& 39.097& --& 19.33& --&-- & --\\
31& 3382& 20:39:15.503& 42:09:10.29& 0.438& 15.946& --& 19.74& --& --& --\\
32& 5823& 20:39:11.660& 42:13:18.73& 0.528& 34.512& --& 19.56& --& --& --\\
33& 7189& 20:38:59.247& 42:29:16.02& 0.136& 2.602& --& 17.52& --& --&-- \\
34& 7743& 20:39:01.818&42:30:56.49 & 0.296& 9.722& --& 18.5& --& --&-- \\
\hline
\end{tabular}
}
\endgroup
\tablefoot{
Sources following the "NXX" naming convention are taken from \citet{Mot07}, those with "DR21-XX" are taken from \citet{Cao19} and "XXXX" sources are from \citet{Cao21}. Where two values are presented, separated by "/", the source was detected in both \citet{Mot07} and \citet{Cao19}. The first value given is from \citet{Mot07} and the second from \citet{Cao19}.
\tablefoottext{a}{Values taken from the catalog where the source was initially reported if available.}
\tablefoottext{b}{Values only available for sources in the \citet{Cao19} and \citet{Cao21} catalogues .}
\tablefoottext{c}{Information only available for sources in the \citet{Cao19} catalogue.}
\tablefoottext{1}{Both DR21-6 and DR21-20 fall within the FWHM of N50.}
}
\end{table*}

\begin{figure*}[t!] 
\centering
    \includegraphics[width=0.3\linewidth]{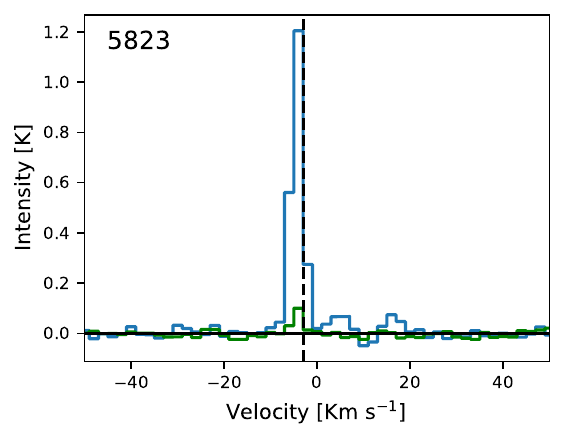}
    \includegraphics[width=0.3\linewidth]{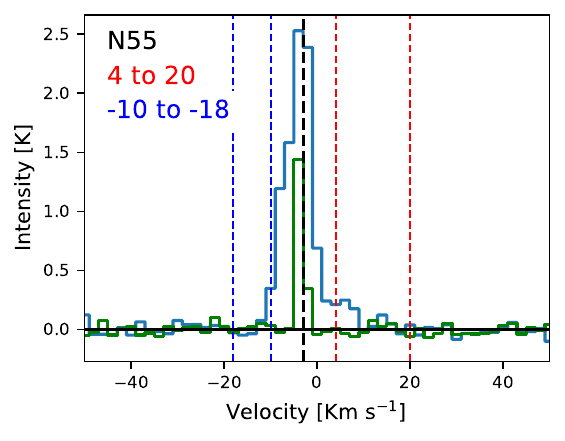}
    \includegraphics[width=0.3\linewidth]{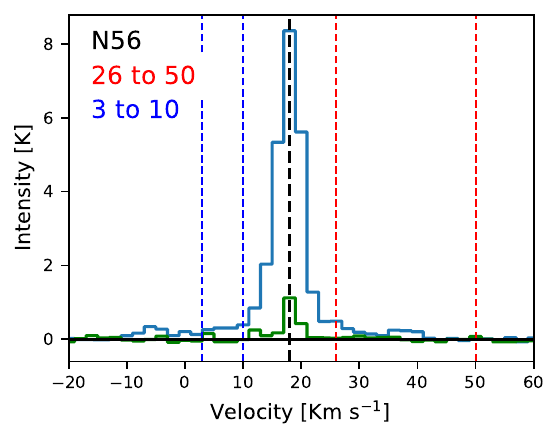}
    \includegraphics[width=0.3\linewidth]{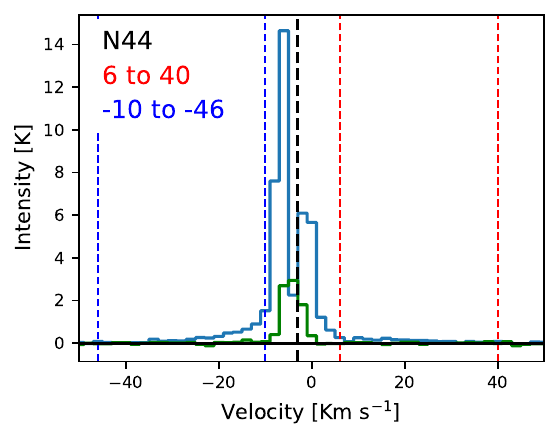}
    \includegraphics[width=0.3\linewidth]{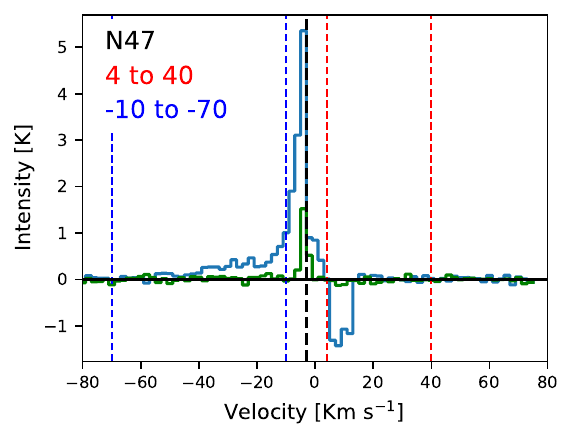}
    \includegraphics[width=0.3\linewidth]{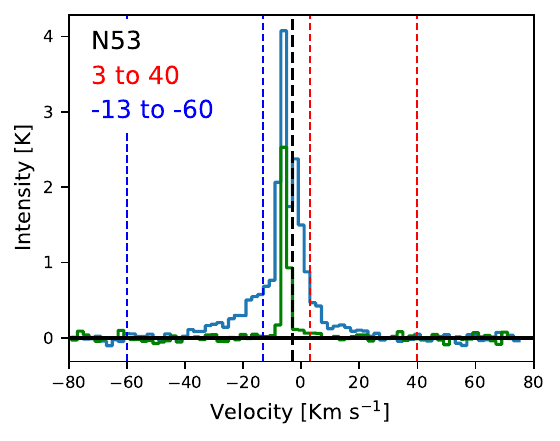}
    \caption{Spectra of HCO$^+$ $J$=1-0 (blue) and H$^{13}$CO$^+$ $J$=1-0 (green) for clumps along DR21 Ridge: 5823, N55, N56, N44, N47 and N53. The spectra were averaged over the FWHM of the corresponding clumps (Table \ref{table:cores}). Red and blue dashed lines mark the velocity range associated with outflow wing emission. The black dashed line marks the source velocity, $v_\text{source}$ = -3 km s$^{-1}$ for all sources except N56, which is located in a separate structure east of DR21, where $V_\text{source}$ = 18 km s$^{-1}$.}
    \label{fig:spectra1}
\end{figure*}

The Cygnus-X region, and especially the DR21 ridge, is a very active high-mass star-forming region, which has already been covered by multiple continuum surveys and there are multiple clump catalogs available in the literature. For this work, we make use of the \cite{Mot07}, \cite{Cao19}, and \cite{Cao21} catalogs, in order to create the most complete sample of clumps in the area of the DR21 ridge, covered by the CASCADE observations.   

We primarily follow the naming conventions from \cite{Mot07}, with sources named as N followed by a number (e.g., \lq N51'). For sources that appear in the \cite{Cao19} catalog and have no corresponding clumps in \cite{Mot07}, we adopt the name from the former publication, which consists of the larger structure name followed by an increasing number (e.g., \lq DR21-12'). Finally, for cores that only appear in the \cite{Cao21} catalog, the four digit number assigned on the core is used (e.g., \lq 5823').

The \cite{Mot07} catalog is based on continuum observations at 1.2 mm, with a resolution of 11$\arcsec$, and using the MAMBO camera of the IRAM 30m telescope. For further details on the extraction of the source catalog we refer to the original publication by \cite{Mot07}.

The \cite{Cao19} catalog makes use of \textit{Herschel} PACS (70 and 160 $\mu$m with resolutions of 9.2\arcsec and 12.6\arcsec respectively), SPIRE (250, 350, and 500 $\mu$m with resolutions of 18.4\arcsec, 25.2\arcsec, and 36.7\arcsec respectively), and JCMT SCUBA-2 (450 and 850 $\mu$m with resolutions of 7.9\arcsec and 13\arcsec respectively) in addition to the aforementioned IRAM 30m (1.2 mm) continuum observations. In addition, we note that only sources with $M > 40 M_\odot$ are included in this catalog.  
The core sample from \cite{Cao21} was derived by applying the $getsources$ algorithm on the $N_\text{H$_2$}$ map, constructed from the SED fits presented in \cite{Cao19}. 
To obtain the most complete census of cores, we included all sources from the \citet{Mot07} catalog, subsequently adding  those from the \citet{Cao19} and \citet{Cao21} catalogs, with no equivalent source reported in \citet{Mot07}.
A summary of all unique cores in the three catalogs, that will be studied in this work, along with their properties, are given in Table \ref{table:cores}. The classification of cores presented here is taken from \citet{Cao19}, where IR-bright refers to cores with mid-IR emission (\textit{Spitzer} 24 $\mu$m and \textit{MSX} 21 $\mu$m) in excess of that of a B3 type stellar embryo ($F_{\textit{Spitzer } 24 \mu \text{m}} \ge$ 23 Jy or $F_{\textit{MSX } 21 \mu \text{m}} \ge$ 17 Jy), IR-quiet to those with emission below that threshold, and starless to those with no mid-IR and no 70 $\mu$m emission.
In addition, the location of all cores is marked in Fig. \ref{fig:DR21_cont}.

\section{Results}
\label{sec:results}
\subsection{Core-averaged spectra}
\label{sec:spectra}

We extracted HCO$^+$ and H$^{13}$CO$^+$ spectra at the location of all sources listed in Table \ref{table:cores}. The spectra are averaged over the area covered by the continuum-measured FWHM of the corresponding source. In cases where FWHM estimates are available from both \citet{Mot07} and \citet{Cao19,Cao21}, the former are used. There are several notable features appearing on spectra of various sources. Some examples of spectra, displaying these features, are shown in Figure \ref{fig:spectra1}. The spectra for all remaining sources are shown in Appendix \href{https://doi.org/10.5281/zenodo.21804190}{G}.

The most common characteristics of the HCO$^+$ spectra are extended line wings (i.e., excess emission seen at velocities larger than the source's $V_\text{lsr}$), suggesting the presence of molecular outflows (see e.g., N47, N53, and N44, Fig. \ref{fig:spectra1}). Depending on the source, the wings extend from $\sim$few km s$^{-1}$ up to $\sim$70  km s$^{-1}$ from the source velocity (indicated by vertical black lines in Fig. \ref{fig:spectra1}). In addition, line wings do not always appear to be symmetrical, with some sources showing a dominant wing (e.g., N47 and N53). Some sources also show signs of absorption at the source velocity that leads to double peaked spectra (e.g., N44 and N53). Since the short spacing information from the IRAM 30m single-dish observations is included in our data, this effect is unlikely to be caused by spatial filtering effects introduced from the interferometric observations. Instead, the double peaked profiles are most likely a result of high optical depth and self-absorption of HCO$^+$.

Two additional velocity components are commonly seen in our sources, apart from the main velocity component of the DR21 ridge at --3 km s$^{-1}$ \citep{Dickel1978,Schneider10}. Channel maps of the HCO$^+$ emission for the three velocity components, located at --3, 9 , and 18 km s$^{-1}$, are shown in Fig. \ref{fig:hco+_components}. The 9 km s$^{-1}$ component is believed to originate from material associated with W75N which extends to the South in front of the DR21 ridge \citep{Dickel1978,Schneider10}. This velocity component is seen in multiple sources along the DR21 ridge, either in absorption (e.g., N47) or emission (e.g., 5823) depending on the level of background emission. 
The 18 km s$^{-1}$ component is seen mostly in sources at the southern end of the DR21 ridge (e.g., 5823). Since this velocity matches well the main velocity component of N56 and N57, it is attributed to extended and diffuse material associated with the structure containing the two sources and located East of the DR21 ridge. 
In addition, certain sources show a \lq shelf'-like emission feature between $\sim$0 and $\sim$10 km s$^{-1}$ (e.g., N55). This is likely the result of the aforementioned 9 km s$^{-1}$ component, shifting to slightly lower velocities, and merging with the emission from the main velocity component.

\subsection{HCO$^{+}$ outflow maps}
\label{sec:outflow_maps}

\begin{figure*}[t!]
\centering
\setlength{\tabcolsep}{0pt}
\renewcommand{\arraystretch}{0}
\begin{tabular}{@{}ccccc@{}}
\hspace{0mm}\includegraphics[width=0.2\textwidth,trim=2.5mm 11mm 1mm 1mm,clip]{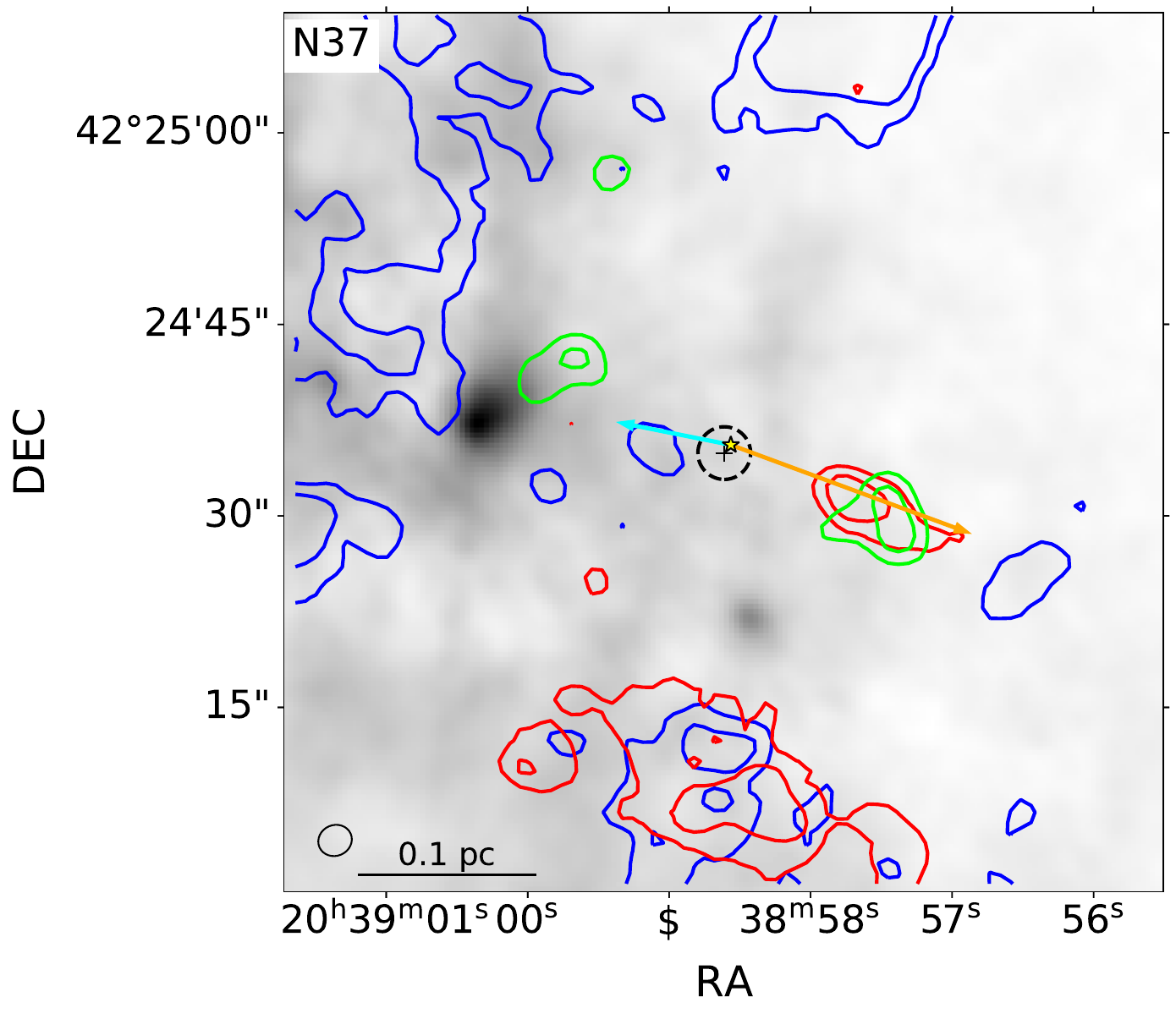} &
\hspace{0mm}\includegraphics[width=0.2\textwidth,trim=2.5mm 11mm 1mm 1mm,clip]{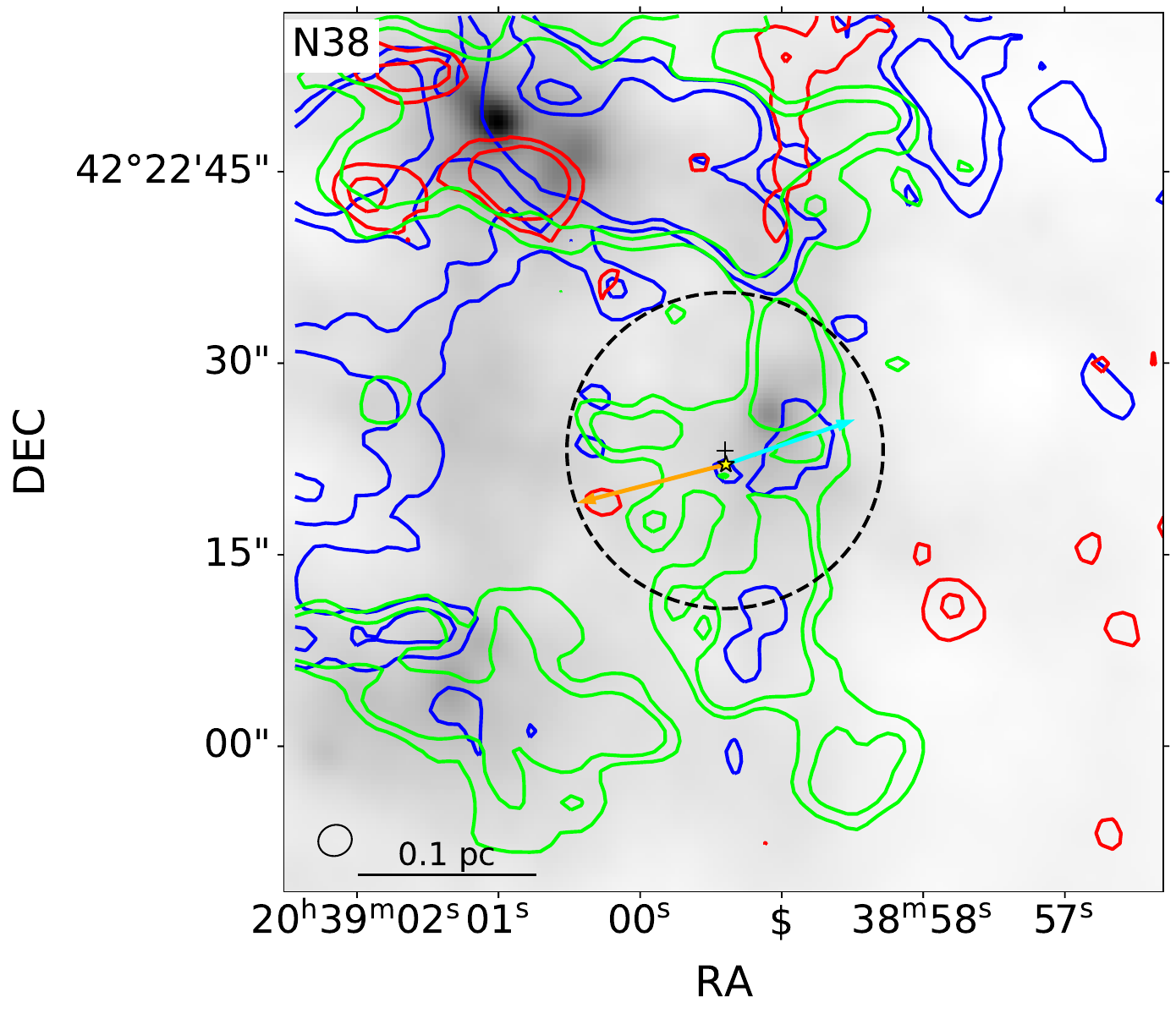} &
\hspace{0mm}\includegraphics[width=0.2\textwidth,trim=2.5mm 11mm 1mm 1mm,clip]{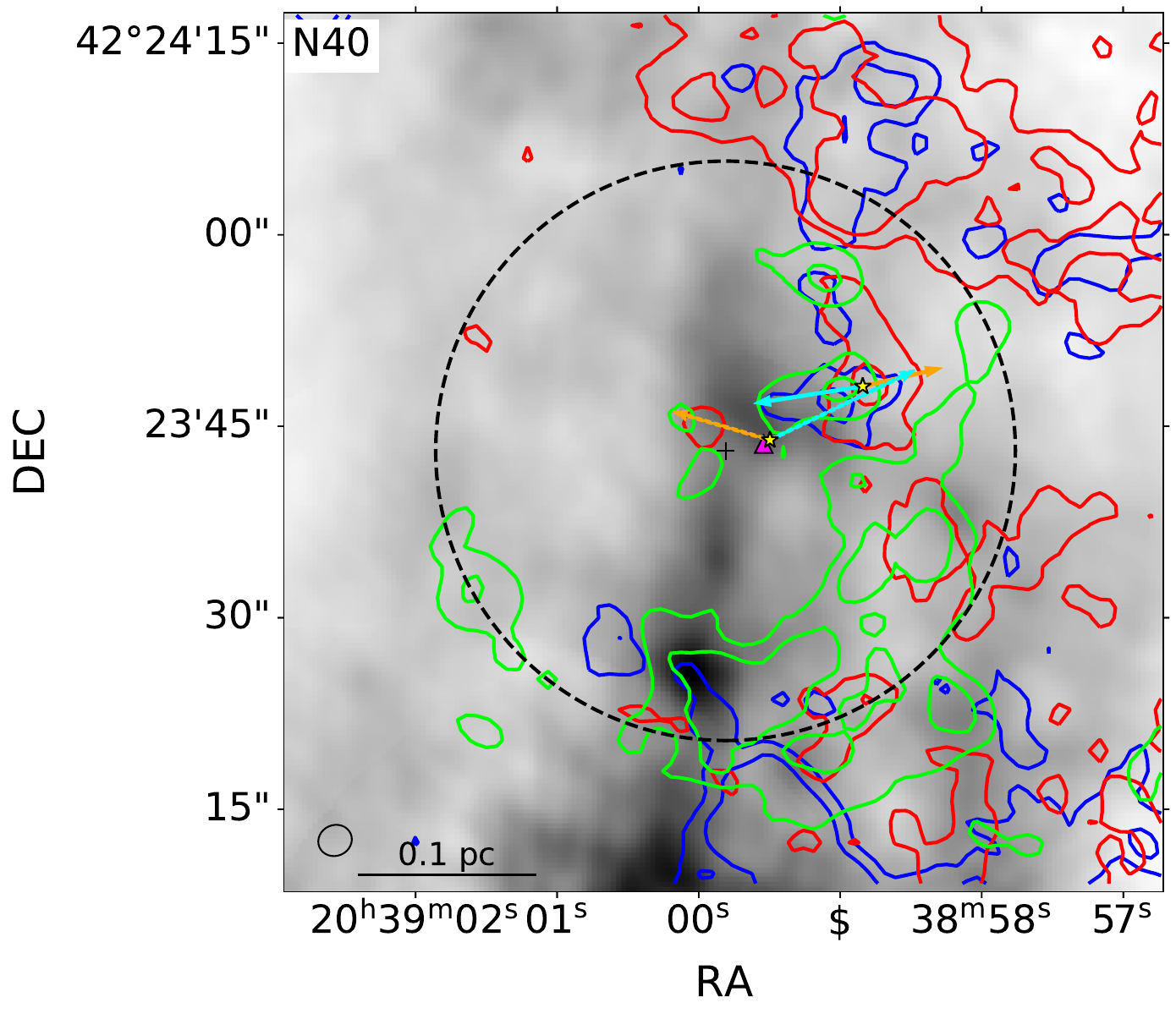} &
\hspace{0mm}\includegraphics[width=0.2\textwidth,trim=2.5mm 11mm 1mm 1mm,clip]{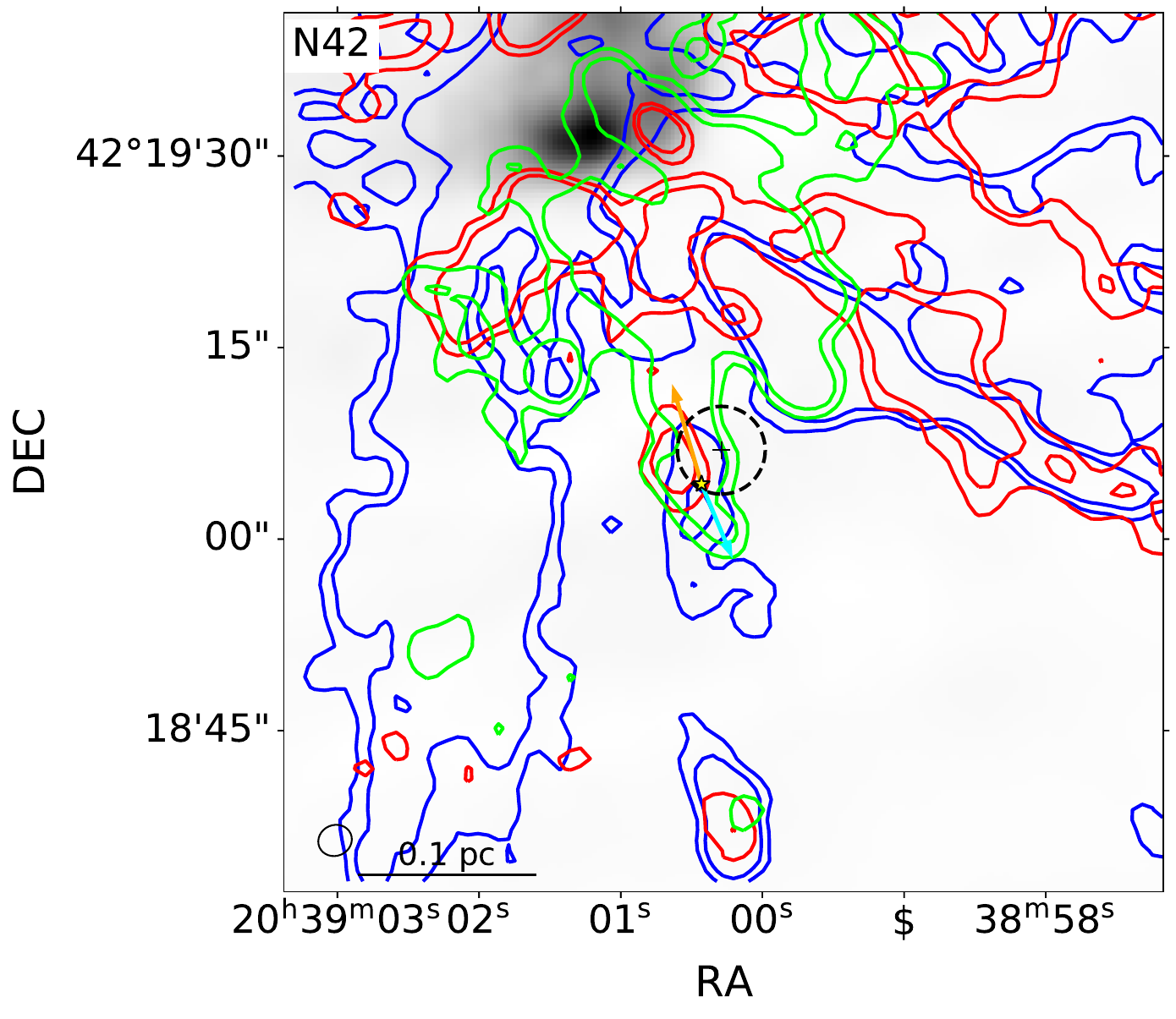} &
\hspace{0mm}\includegraphics[width=0.21\textwidth,trim=2.5mm 11mm 5mm 2mm,clip]{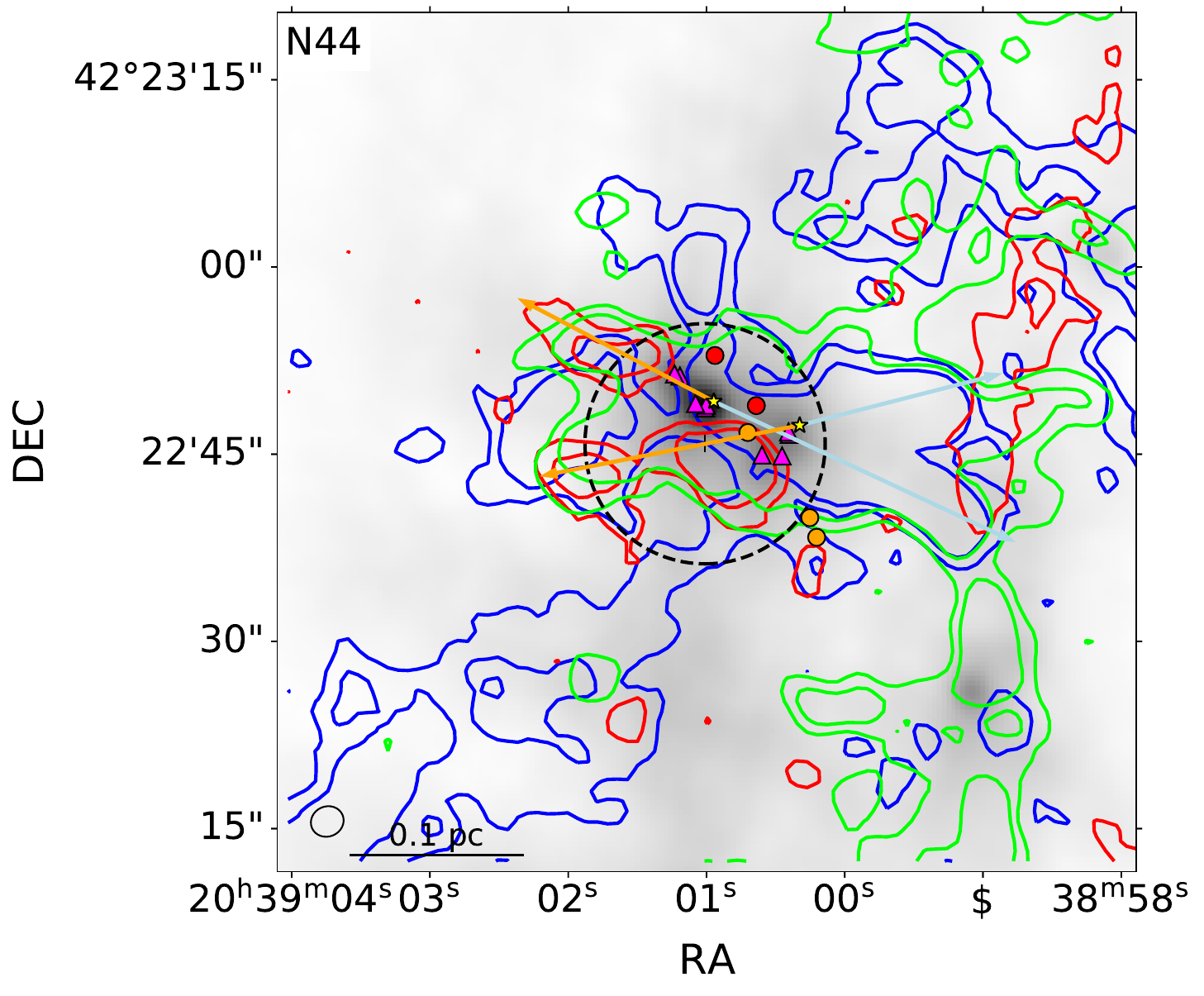} \\[-1pt]

\hspace{0mm}\includegraphics[width=0.21\textwidth,trim=2.5mm 11mm 5mm 2mm,clip]{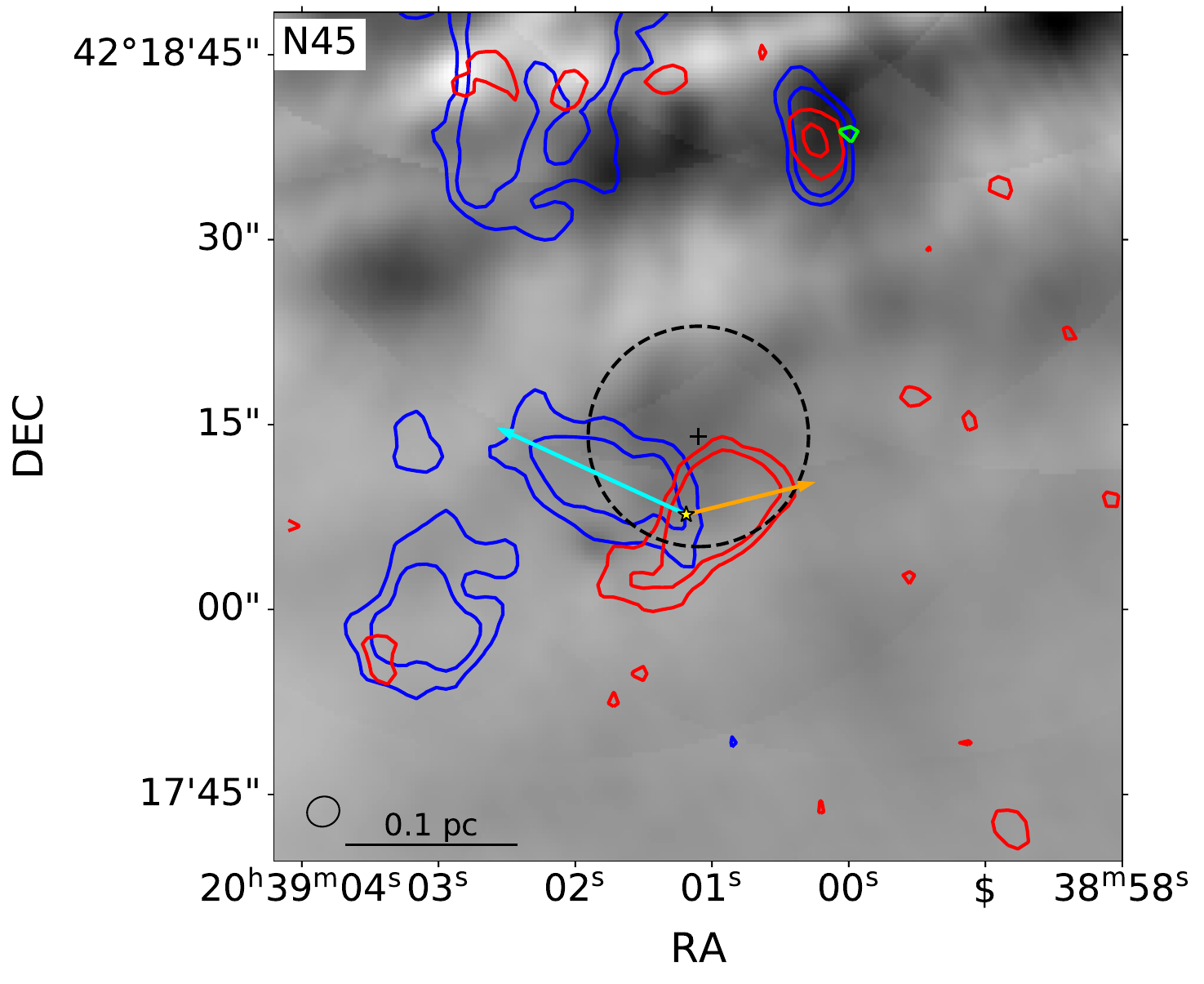} & 
\includegraphics[width=0.2\textwidth,trim=2.5mm 11mm 1mm 1mm,clip]{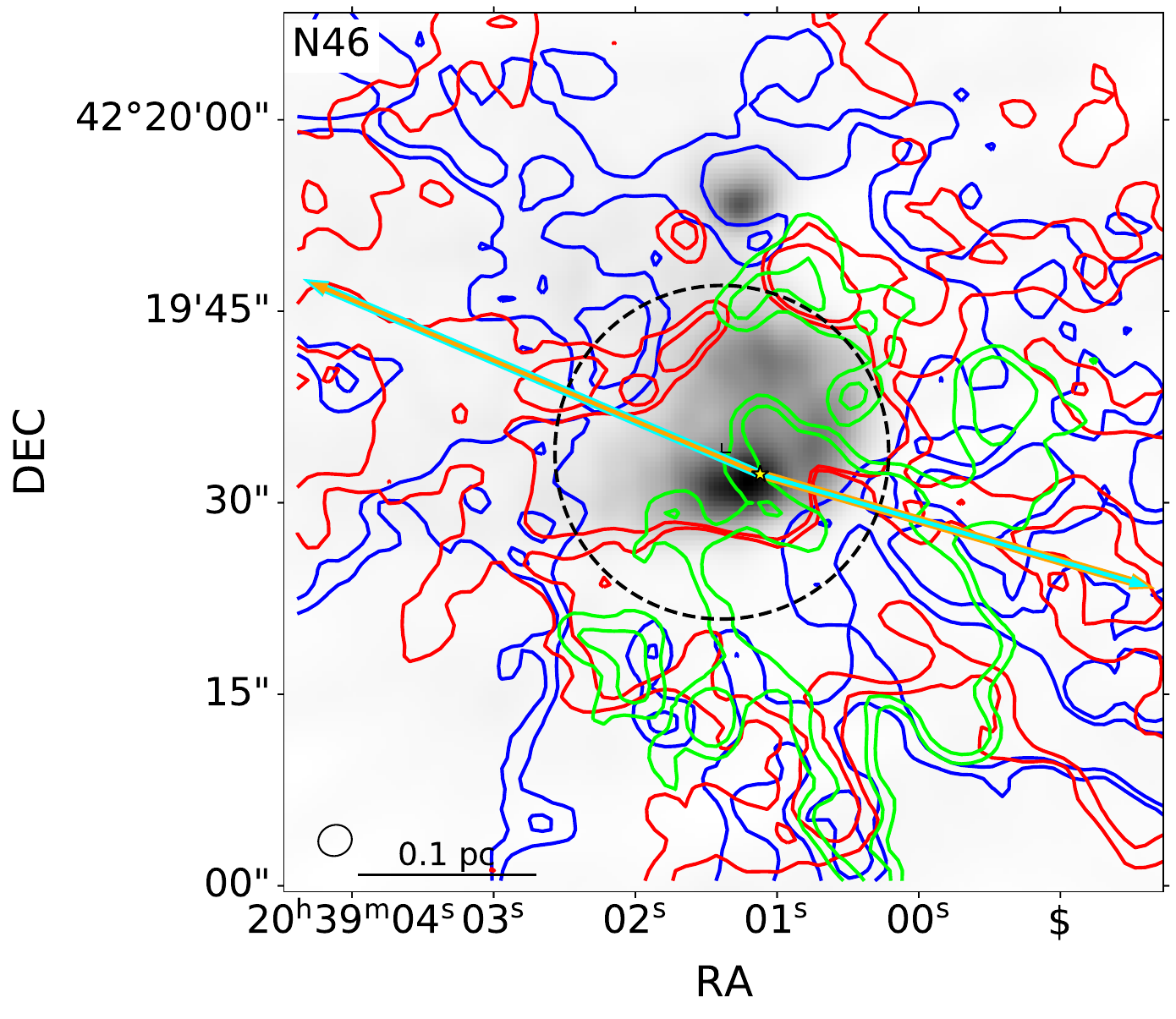} &
\hspace{0mm}\includegraphics[width=0.2\textwidth,trim=2.5mm 11mm 1mm 1mm,clip]{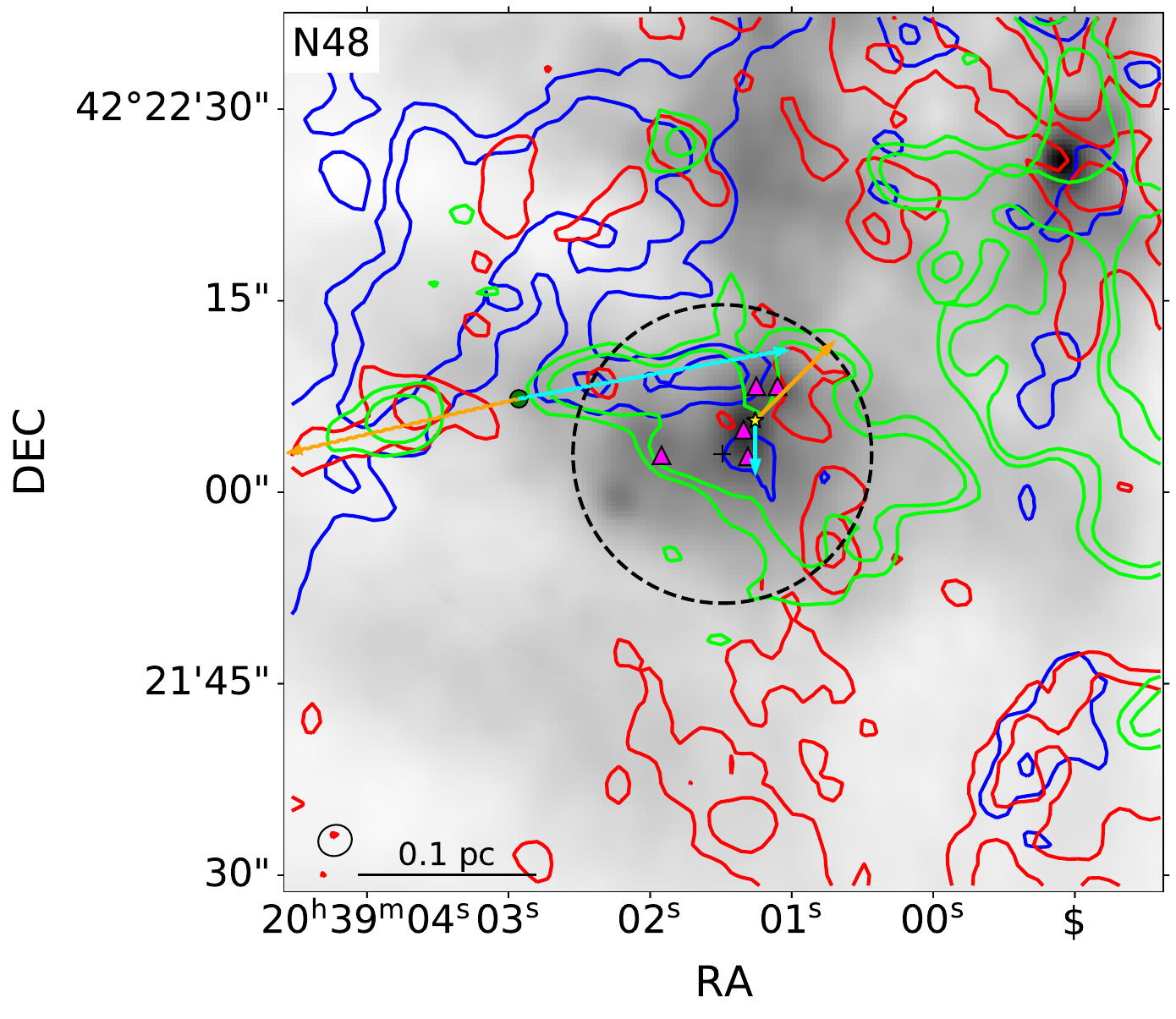} &
\hspace{0mm}\includegraphics[width=0.2\textwidth,trim=2.5mm 11mm 1mm 1mm,clip]{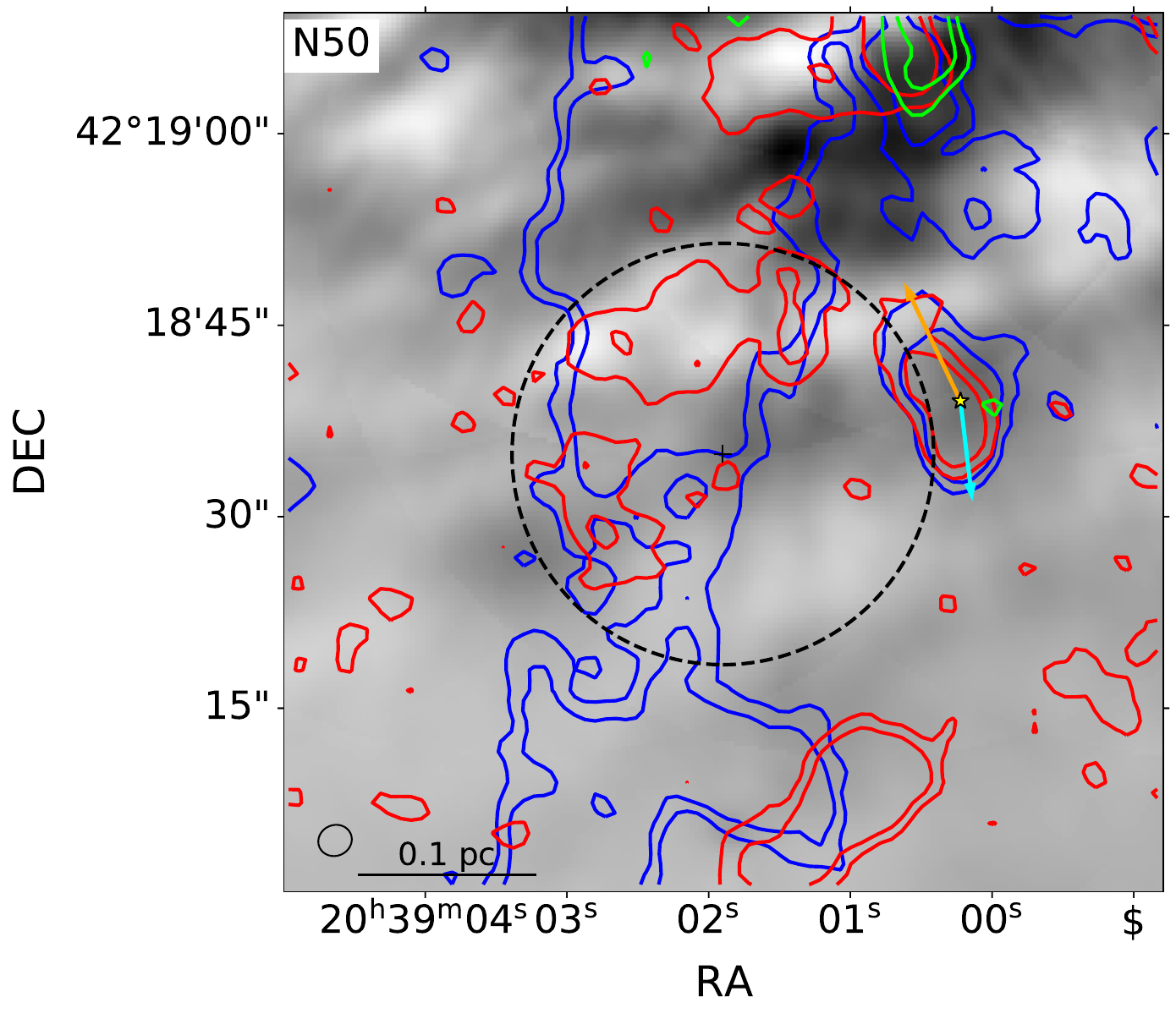} &
\includegraphics[width=0.2\textwidth,trim=2.5mm 3mm 0mm 0mm,clip]{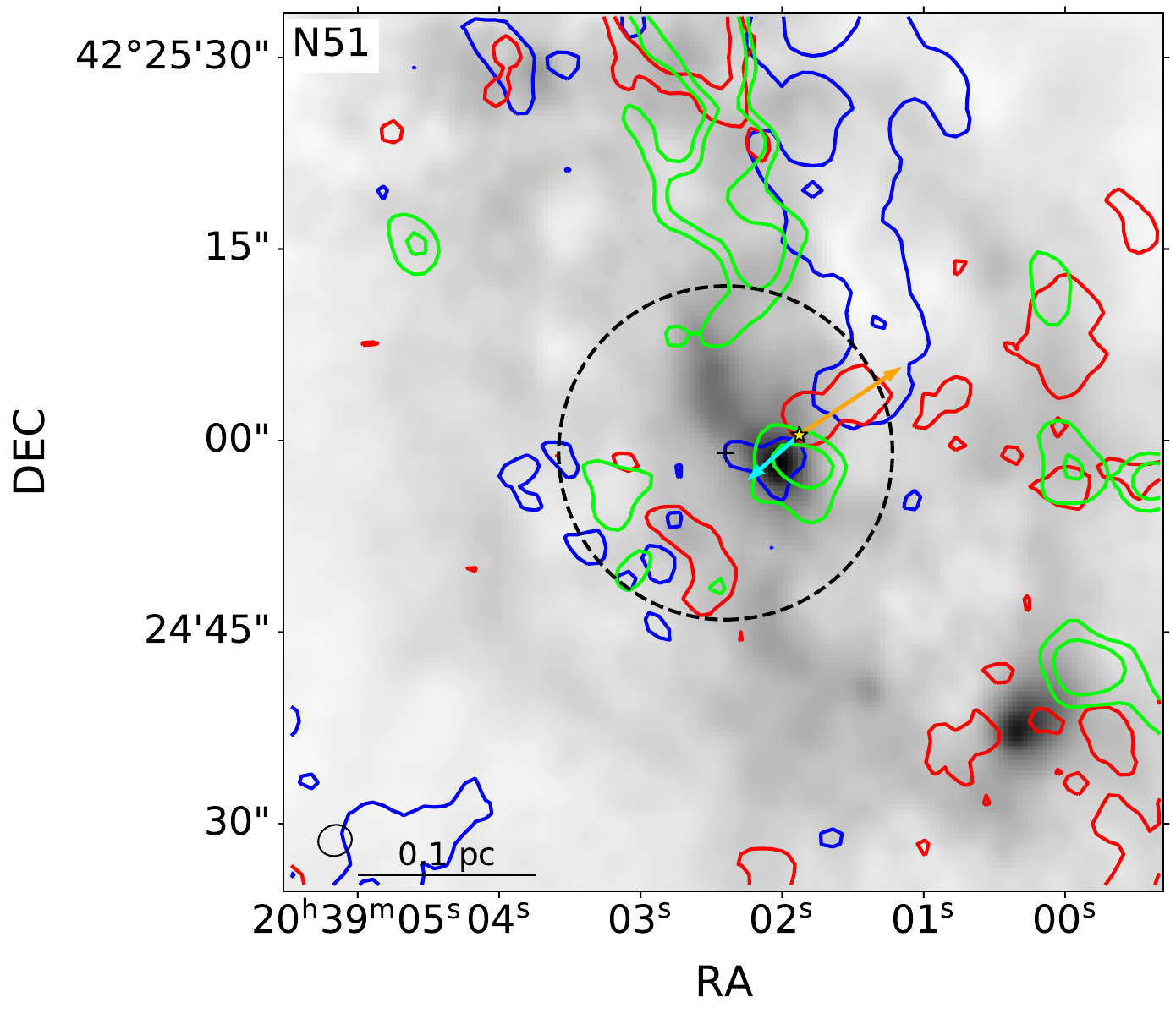} \\[-1pt]

\hspace{0mm}\includegraphics[width=0.2\textwidth,trim=2.5mm 3mm 1mm 1mm,clip]{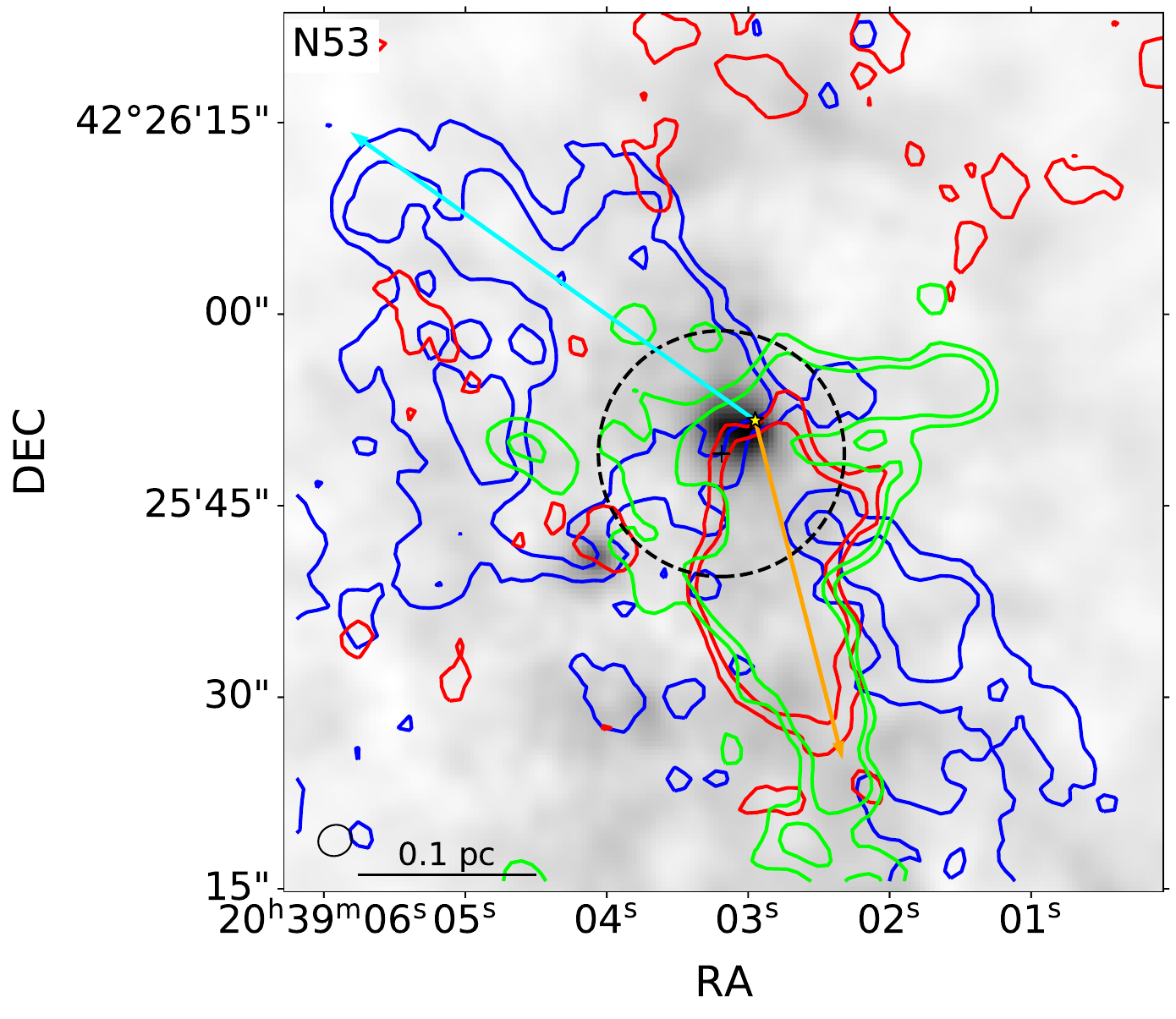} &
\hspace{0mm}\includegraphics[width=0.2\textwidth,trim=2.5mm 3mm 1mm 1mm,clip]{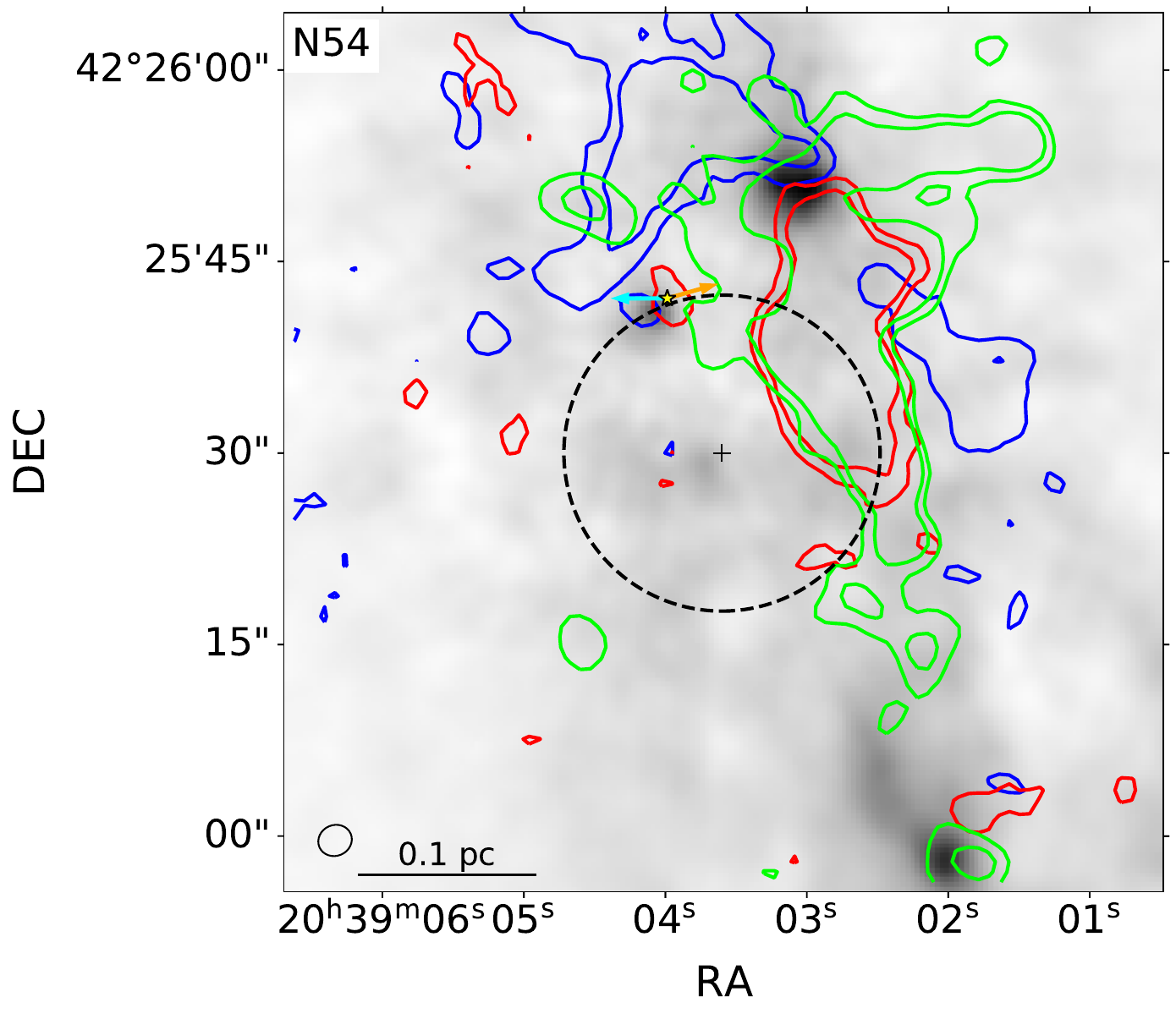} &
\includegraphics[width=0.2\textwidth,trim=2.5mm 3mm 1mm 0mm,clip]{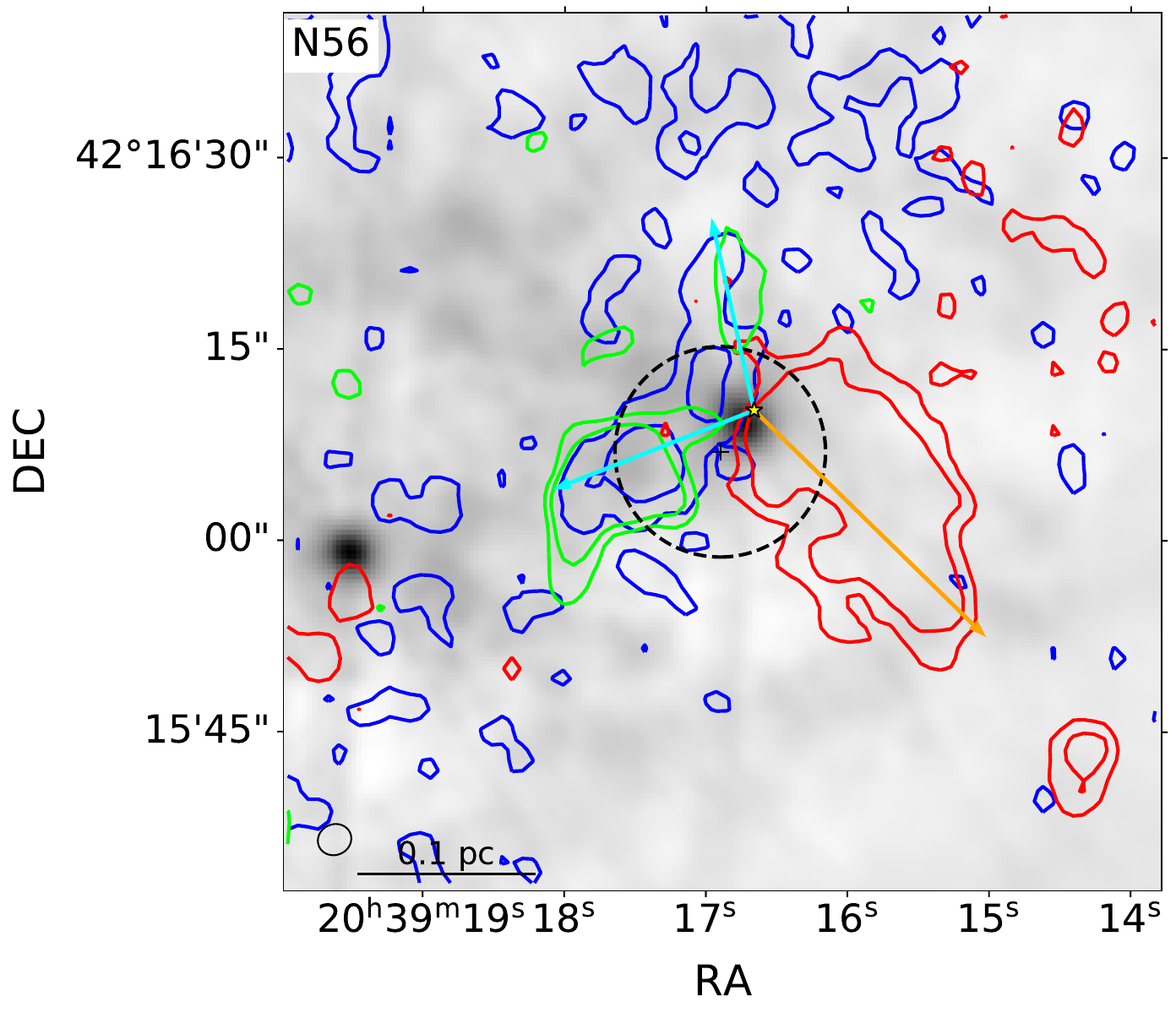} &
\hspace{0mm}\includegraphics[width=0.2\textwidth,trim=2.5mm 3mm 1mm 0mm,clip]{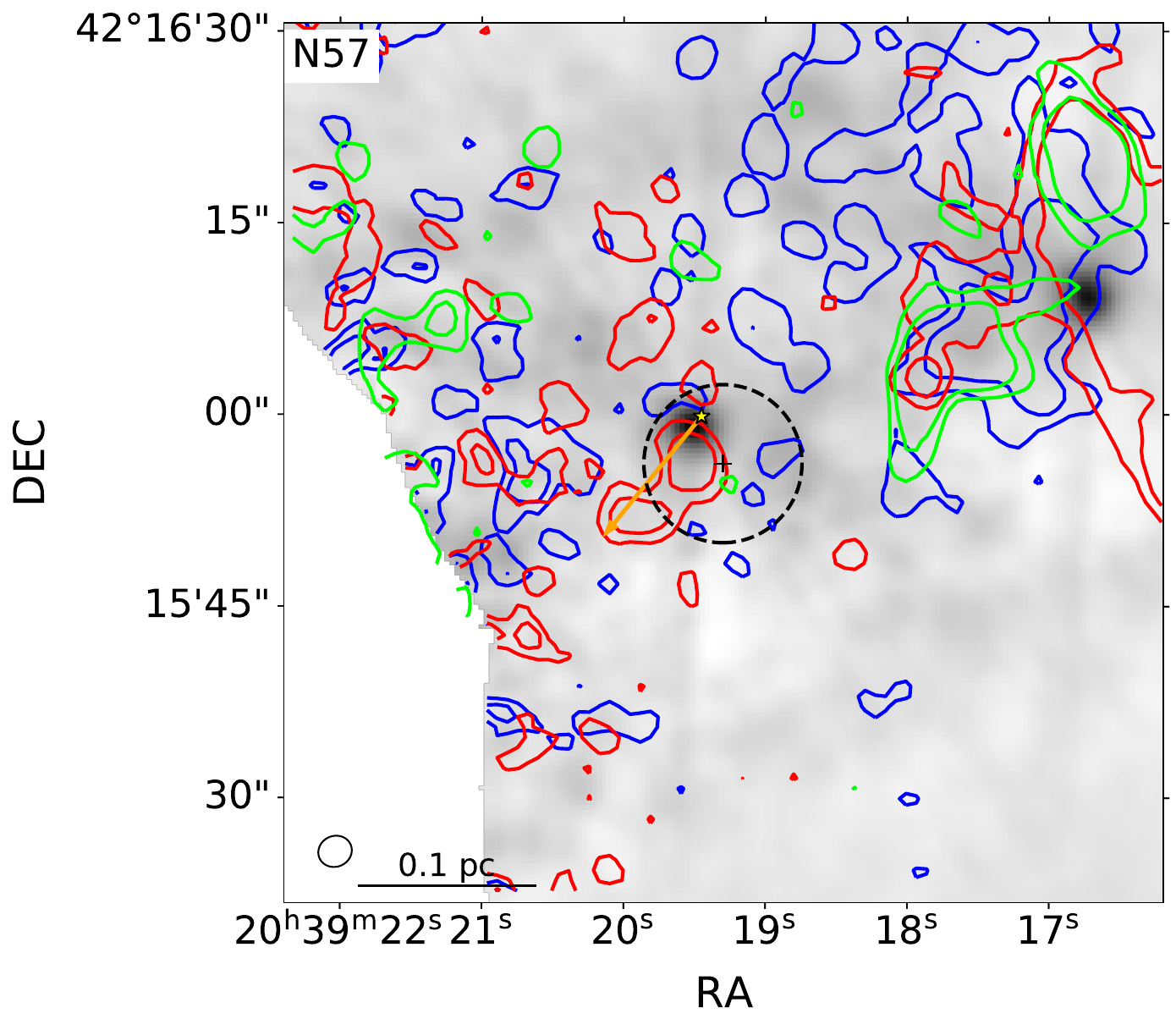}
\\
\end{tabular}
\caption{Velocity-integrated maps of the red-shifted (red contours) and blue-shifted (blue contours) HCO$^+$ emission. Velocity ranges used for each source are given in Table \ref{table:outflow_velocities}. Green contours show the integrated SiO intensity. The black cross and dashed circle indicate the location and size of the corresponding core (see Table \ref{table:cores}). Red and blue arrows mark the most likely direction of the outflow, and the yellow star indicates the outflow origin. Finally, the grayscale background shows the 3.6 mm continuum from NOEMA and GBT.}
\label{fig:outflowmaps1}
\end{figure*}

As mentioned in Section \ref{sec:spectra}, a large number of sources in our sample show substantial wing features in their core-averaged spectra. Such wings are often signs of ongoing outflow activity, originating from the protostars within the cores \citep[e.g.,][]{Yang2022}. 
Identifying protostellar outflows in regions such as the DR21 ridge is not a trivial process. The high density of star formation in the region means that multiple outflows can appear in close proximity, creating confusion regarding the exact origin of each individual high-velocity component. Adding to the complexity are outflows from more evolved sources than the ones examined in this work, as well as large-scale flows along the filament that can appear as either blue- or red-shifted \citep{Schneider10}.

We examined in detail integrated intensity maps of the high-velocity HCO$^+$ emission as well as channel maps of the same tracer. In addition, we used the 3 mm CASCADE continuum, along with available literature information, in order to connect the detected high-velocity HCO$^+$ components to either a continuum driving source, or to a previously reported molecular outflow.
Figure \ref{fig:outflowmaps1} shows the HCO$^+$ outflow emission for sources with confirmed outflows. Maps for sources with velocity wings or other notable features seen in their spectra, such as emission in additional velocity components or a \lq shelf'-like emission feature (see Sec. \ref{sec:spectra}), but no clear outflow, are shown in Fig. \href{https://doi.org/10.5281/zenodo.21804190}{D.1}. The velocity ranges used for the contours in each source were selected by comparing, by eye, the HCO$^+$ spectrum with that of the usually optically thin H$^{13}$CO$^+$ isotopologue. Under the assumption that H$^{13}$CO$^+$ probes primarily the dense core, HCO$^+$ emission seen extending at higher velocities than those of H$^{13}$CO$^+$ is attributed to outflow emission.
The velocity ranges used for each source are summarized in Table \ref{table:outflow_velocities}. For sources in the DR21 ridge, the channel at 9 km s$^{-1}$ is not included in the integration in order to avoid contamination from the foreground component.  
We discuss each individual outflow source in more detail in Appendix \ref{app:outflowsources}.

\begin{table}
\caption{Outflow velocity ranges in units of km s$^{-1}$} 
\label{table:outflow_velocities} 
\centering
\small
\begin{tabular}{l c c c c c}
\hline \hline 
Name & \multicolumn{2}{c}{Blue} & \multicolumn{2}{c}{Red} \\  
\hline
 & $v_\text{in}$ & $v_\text{out}$ & $v_\text{in}$ & $v_\text{out}$  \\
\hline
N38 &-13&-25&13&25\\
N42 &-16&-24&12&22  \\
N45 &-17&-35&10&30\\
N48 &-12&-25&2&10  \\
N50 &-8&-50&4&24 \\
N53 &-13&-60&3&40 \\
N54 &-22&-55&4&40 \\
N37 &-10 &-20 & 12& 30\\
N40 &-10 &-22 &6 &25 \\
N51 & -18&-30 & 4& 26\\
N56 & 3& 10& 26& 50\\
N44 & -10&-46 & 6&40 \\
N57 & --& --& 23&30\\
\hline
\end{tabular}

\end{table}

Overall, the analysis of the HCO$^+$ emission in the vicinity of all 34 dense molecular clumps reported in the DR21 ridge revealed 14 molecular outflows, a detection rate of $\sim$40~\%. We note though that our analysis cannot detect outflows extending along the plane of the sky, thus the true number of outflows can potentially be higher. The number of such outflows though is presumed to be only a small percentage of the total number of outflows in the region, and therefore their exclusion from the subsequent analysis is not expected to substantially impact the conclusions. In the vast majority of outflow sources, both lobes are detected, with just a single source displaying only red-shifted emission. Constraining the outflow emission enables us to further advance the analysis by estimating the energetic properties of the outflows.

\subsection{Outflow properties}
\label{sec:outflow_props}

The HCO$^+$ observations of CASCADE allow for the calculation of the typical outflow properties such as the outflow force, $F$, the outflow mass, $M$, and the kinetic energy, $E_\text{kin}$ for all the detected outflows discussed in Section \ref{sec:outflow_maps}. 
For the calculation we follow the same approach as in \cite{Skretas23}, where the properties of the outflow of DR21 Main were estimated. 
\begin{table}
\caption{Inclination angle correction factors for the calculation of outflow forces using the Separation method}
\label{table:inclinations} 
\centering
\small
\begin{tabular}{c c c c c c} 
\hline\hline 
$i$\tablefootmark{a} ($\degr$) & 10 & 30 & 50 & 70 \\ 
\hline
$c_3$\tablefootmark{b} & 0.6 & 1.3 & 2.4 & 3.8 & \\
\hline 
\end{tabular}
\tablefoot{
\tablefoottext{a}{$i$ is measured from the line of sight.}
\tablefoottext{b}{Values are interpolated from Table 6 of \citet{Downes2007}}, where $\alpha = 90 - i$. 
}
\end{table}

In brief, the outflow force is determined using the Separation method presented in \cite{vdm13}. In this method, the outflow force for a single outflow lobe is given by 
\begin{equation}
    F_{\text{HCO}^+} = c_3 \times \frac{\displaystyle K \left( \sum_\text{j} \left[ \int_{v_\text{in}}^{v_\text{out,j}}T(v')v'\text{d}v' \right]_\text{j}\right) v_\text{max}}{\displaystyle R_\text{lobe}} .
    \label{eq:outflow force}
\end{equation}
Here, $c_3$ is a correction factor depending on the inclination angle of the outflow (Table \ref{table:inclinations}), $K$ is a conversion factor between the line integrated intensity and the molecular gas mass, and the integral $\int_{v_\text{in}}^{v_\text{out,j}}T(v')v'\text{d}v'$ is the velocity weighted integrated intensity. In addition, $v_\text{max}$ corresponds to the maximum line-of-sight velocity in the outflow lobe, and $R_\text{lobe}$ is the length of the outflow lobe, while the sum runs over all pixels (j) that are part of the outflow. In turn, the line intensity-to-mass conversion factor $K$ is given by
\begin{equation}
    K = \mu m_\text{H} A \frac{8 \pi k_\text{B}\nu^2}{h c^3 A_\text{ul}} \left[ \frac{\text{H}_2}{\text{HCO$^+$}}\right] \frac{Q{(T_\text{exc})}}{g_\text{u}}\text{e}^{E_\text{u}/T_\text{exc}}
    \label{eq:kappa}
\end{equation}
where $\mu$ is the mean molecular weight ($\mu$=2.8 used in this work), $m_\text{H}$ is the hydrogen mass, $A$ is the observed area of the outflow, $[\text{H}_{2}/\text{HCO}^+]$ is the abundance ratio between H$_2$ and HCO$^+$, $Q{(T_\text{exc})}$ is the partition function at a specific excitation temperature $T_\text{exc}$, $g_\text{u}$ is the degeneracy of the upper level of the observed transition, $E_\text{u}$ is the upper level energy in Kelvins, and $\nu$ is the frequency of the observed transition in Hz. Additionally, $c$ is the speed of light, $k_\text{B}$ is Boltzmann's constant, $h$ is Planck's constant and $A_\text{ul}$ is Einstein A coefficient for the transition in s$^{-1}$. A more detailed description of the conversion factor is presented in Appendix C of \cite{vdm13}.
In this work, we assume a single excitation temperature of 40 K for all sources, which was calculated for DR21 Main \citep{Gar91}, while we also adopt the abundance ratio of H$_2$ over HCO$^+$ of $1.6\times10^{8}$, also estimated for DR21 Main \citep{Gar92} and similar to what was used in \cite{Skretas23}. 
We discuss the impact of these assumptions on the resulting outflow properties later in this Section. The corresponding value of the partition function and the remaining molecular data are taken from the CDMS line database \citep{Muller2001}.

For the remaining outflow properties, the total gas mass carried by the outflow is calculated  as:
\begin{equation}
    M_\text{out} = K \left( \sum_\text{j} \left[ \int_{v_\text{in}}^{v_\text{out,j}}T(v')\text{d}v'\right]_\text{j}\right), 
\end{equation}
while, the time-averaged kinetic energy of the outflow is calculated as:
\begin{equation}
E_\text{kin} = \frac{1}{2}M_\text{out} \langle v \rangle^2 ,
\end{equation}
with $\langle v \rangle$ the average value of the maximum outflow velocity detected in each pixel of the outflow lobe, $v_\text{out,j}$.
Additionally, the momentum as:
\begin{equation}
    P = \displaystyle K \left( \sum_\text{j} \left[ \int_{v_\text{in}}^{v_\text{out,j}}T(v')v'\text{d}v' \right]_\text{j}\right).
    \label{eq:momentum}
\end{equation}

The dynamical time, representing a rough estimate of the age of each outflow, can be calculated as:
\begin{equation}
    t_\text{dyn} = \frac{R_\text{lobe}}{v_\text{max}},
\end{equation}

\begin{table*}
\caption{Outflow properties} 
\label{table:outflow_properties} 
\centering 
\small
\begingroup
\makebox[\textwidth][c]{
\begin{tabular}{l c c c c c c c c c}
\hline \hline 
Name & & \begin{tabular}[c]{@{}c@{}}$R_\text{lobe}$\\  {[}pc{]}\end{tabular}& \begin{tabular}[c]{@{}c@{}}$t_\text{dyn}$\\  {[}yr{]}\end{tabular} & \begin{tabular}[c]{@{}c@{}}$M_\text{out}$\\  {[}$\times10^{-2}$ M$_\odot${]}\end{tabular} & \begin{tabular}[c]{@{}c@{}}$p_\text{out}$\\  {[}M$_\odot$ km s$^{-1}${]}\end{tabular} & \begin{tabular}[c]{@{}c@{}}$E_\text{kin}$\\  {[}$\times10^{43}$ erg{]}\end{tabular}& \begin{tabular} [c]{@{}c@{}}$ \dot{M}_\text{out}$\\  {[}$\times10^{-5}$ M$_\odot$ yr$^{-1}${]}\end{tabular} & \begin{tabular}[c]{@{}c@{}}$F_\text{out}$\\  {[}$\times10^{-4}$ M$_\odot$ km yr$^{-1}$ s$^{-1}${]}\end{tabular} & \begin{tabular}[c]{@{}c@{}}$L_\text{kin}$\\  {[}$\times10^{40}$ erg yr$^{-1}${]}\end{tabular}  \\  
\hline
N38 & Blue & 0.06 & 1000 & 1.4 & 0.25 & 3.3 & 2.2 & 3.7 & 5.0 \\
 & Red & 0.07 & 920 & 1.0 & 0.19 & 4.5 & 1.7 & 3.2 & 7.4 \\
 & Total & -- & -- & 2.4 & 0.44 & 7.8 & 3.8 & 6.9 & 12.0 \\
N42 & Blue & 0.07 & 950 & 2.6 & 0.54 & 14.0 & 4.2 & 8.6 & 23.0 \\
 & Red  & 0.05 & 580 & 9.4 & 1.63 & 42.0 & 24.2 & 42.1 & 110.0 \\
 & Total& -- & -- & 12.0 & 2.17 & 56.0 & 28.4 & 50.7 & 133.0 \\
N45 &  Blue& 0.10 & 700 & 20.8 & 4.75 & 150.0 & 44.7 & 102.0 & 330.0 \\
 & Red & 0.06 & 470 & 38.6 & 8.37 & 180.0 & 123.8 & 268.7 & 57.0 \\
 & Total & -- & -- & 59.4 & 13.12 & 330.0 & 168.5 & 370.7 & 387.0 \\
N48 & Blue& 0.02 & 410 & 0.2 & 0.02 & 0.3 & 0.6 & 0.9 & 1.2\\
 & Red& 0.04 & 820 & 8.0 & 0.45 & 7.4 & 14.8 & 8.4 & 14.0 \\
 & Total& -- & -- & 8.2 & 0.47 & 7.7 & 15.4 & 9.3 & 15.2 \\
N50 & Blue& 0.04 & 380 & 36.5 & 5.57 & 210.0 & 143.3 & 218.5 & 810.0\\
 & Red & 0.06 & 880 & 32.6 & 3.05 & 90.0 & 55.5 & 52.1 & 150.0 \\
 & Total& -- & -- & 69.1 & 8.72 & 300.0 & 198.8 & 270.6 & 960.0 \\
N53 & Blue& 0.25 & 1520 & 171.6 & 37.62 & 1300.0 & 170.7 & 374.1 & 1300.0 \\
 & Red& 0.17 & 1070 & 227.3 & 33.94 & 1500.0 & 320.1 & 478.1 & 2100.0 \\
 & Total& -- & -- & 398.9 & 71.56 & 2800.0 & 490.8 & 852.2 & 3400.0 \\
N54 & Blue& 0.02 & 210 & 3.5 & 0.93 & 30.0 & 25.1 & 67.4 & 220.0\\
 & Red & 0.02 & 340 & 3.4 & 0.29 & 5.5 & 15.1 & 12.8 & 24.0\\
 & Total & -- & -- & 6.9 & 1.22 & 35.5 & 40.2 & 80.2 & 244.0 \\
N37 & Blue& 0.05 & 970 & 0.5 & 0.07 & 1.2 & 0.8 & 1.0 & 1.8\\
 & Red& 0.13 & 1170 & 12.4 & 2.42 & 59.0 & 16.1 & 31.3 & 76.0 \\
 & Total&-- & -- &  12.9& 2.49 & 60.2 & 16.9 & 32.3 & 77.8 \\
N40 & Blue & 0.08 & 1650 & 0.4 & 0.05 & 1.0 & 0.4 & 0.5 & 0.9 \\
 & Red & 0.05 & 670 & 1.0 & 0.12 & 1.0 & 2.2 & 2.6 & 2.2 \\
 & Total& -- & -- & 1.4 & 0.17 & 2.0 & 2.6 & 3.1 & 3.1 \\
N51 & Blue & 0.03 & 350 & 0.3 & 0.07 & 1.7 & 1.5 & 3.2 & 7.2 \\
 & Red & 0.06 & 780 & 10.2 & 0.93 & 19.0 & 19.7 & 17.9 & 38.0 \\
 & Total& -- & -- & 10.5 & 1.00 & 20.7 & 21.2 & 21.1 & 45.2 \\
N56 & Blue& 0.11 & 940 & 21.0 & 3.76 & 100.0 & 33.7 & 60.4 & 160.0\\
 & Red& 0.16 & 1310 & 168.9 & 23.04 & 580.0 & 194.7 & 265.6 & 660.0 \\
 & Total & -- & -- & 189.9 & 26.80 & 680.0 & 228.4 & 326.0 & 820.0 \\
N44 & Blue & 0.17 & 980 & 137.0 & 31.55 & 930.0 & 211.9 & 487.9 & 1400.0 \\
 & Red& 0.11 & 610 & 77.6 & 13.68 & 240.0 & 193.2 & 340.4 & 590.0 \\
 & Total & -- & -- & 214.6 & 45.23 & 1170.0 & 405.1 & 828.3 & 1990.0 \\
N57 & Blue& -- & -- & -- & -- & -- & -- & -- & -- \\
 & Red & 0.07 & 1440 & 9.7 & 0.74 & 9.7 & 10.1 & 7.8 & 10.0 \\
 & Total& -- & -- & -- & -- & -- & -- & -- & -- \\
\hline
\end{tabular}
}
\endgroup

\end{table*}

\noindent allowing, in turn, the calculation of the mass loss rate:
\begin{equation}
    \dot{M}_\text{out} = \frac{M_\text{out}}{t_\text{dyn}},
\end{equation}
and the kinetic power of the outflows:
\begin{equation}
    L_\text{kin} = \frac{E_\text{kin}}{t_\text{dyn}}.
\end{equation} 

The outflow properties are determined separately for the blue- and red-shifted outflow lobes, and the resulting properties are listed in Table \ref{table:outflow_properties}. 

When we compare the CASCADE HCO$^+$ results with those from SMA presented in \citet{Skretas22}, we find the CASCADE outflow forces to be $\sim 1-2$ orders of magnitude larger for all sources that are common to both studies (N54, N53, N51, N48, and N38).
The differences might be due to the use of a higher excitation transition of a different molecular tracer (CO $J=3-2$) meaning that the two studies are probing slightly different outflow components. At the same time, the lack of short spacings in the SMA observations can also lead to an underestimate of the outflow properties. 
However, CO $J=2-1$ from the Plateau de Bure Interferometer (PdBI) observations from \citet{duarte13}, which included short spacings from the IRAM 30~m telescope, still appear lower than our HCO$^+$ results by up to an order of magnitude. In contrast, the estimates of the outflow mass, momentum and kinetic energy for the outflow of N56, presented in \citet{Deb2021}, using observations of CO ($J=3-2$) taken with the JCMT appear in good agreement with those estimated in this work for the same source.
In Section \ref{sec:covshco+}, we expand further on the potential reasons of the apparent differences between outflow properties estimated using CO and HCO$^+$ observations. 

To determine the outflow properties, we adopted a number of assumptions as follows. Firstly, in order for the conversion factor $K$ to be accurate, optically thin emission is required. While the ridge displays some very high column densities, the optically thin assumption is in general believed to hold for the higher velocities in protostellar outflows \citep{Gar92}. This is found to be true even in the massive outflow of DR21 Main \citep[][]{Gar92,Skretas23}, located in the densest region along the ridge.
To avoid including material from the dense ridge in the calculation, we excluded channels with velocities close to that of the main filament (-3 km s$^{-1}$, also see Table \ref{table:outflow_velocities} for the exact velocity ranges used in each source). 
This has the potential to lead to an underestimate of the measured mass, with \citet{Offner2011} suggesting, based on simulations, that integrating over velocities >2 km s$^{-1}$ can lead to an underestimate of the outflow mass by a factor of 5–10. The impact on $F$ or $E_\text{kin}$ is not as significant due to the dependence of those parameters on $v^2$, and thus to the higher velocity material. Also, in order to avoid contamination from the foreground component at 9 km s$^{-1}$ the corresponding channel is excluded from the calculation of the outflow properties. This will lead to a slight underestimate of the true outflow properties, but we expect the difference to be relatively low and without significant impact on our conclusions. The velocity component at $\sim$18 km s$^{-1}$ is not seen in absorption towards any of the outflow sources.
Secondly, the correction factor of inclination angle, $c_3$, has been calculated only for the inclination angles of 10$\degr$, 30$\degr$, 50$\degr$, and 70$\degr$ (Table \ref{table:inclinations}). As the inclination of the outflows is not trivial to estimate, in this work the outflow properties were estimated using all four different factors and the average values are provided in Table \ref{table:outflow_properties}. Depending on the real inclination of each source, the true values can differ from the assumed average by a factor of up to $\sim$6.
Additionally, a uniform excitation temperature of 40 K was assumed for all outflows along the ridge, the same as that found by \citet{Gar91} for the outflow of DR21 Main, and used in \citet{Skretas23}. While this temperature might not be very accurate for all outflows in our sample it has been shown that changes in excitation temperatures of an order of a few Kelvins have small impact on the resulting outflow properties, with \citet{Skretas22} showing that an increase of the excitation temperature from 50\,K to 100\,K leads to an increase of the outflow force only by a factor of 1.4.
Finally, all outflow parameters estimated here depend linearly on the assumed H$_2$ over HCO$^+$ ratio. 
The adopted H$_2$/HCO$^+$ ratio of 1.6 $\times 10^{8}$ lies within the range of values reported for high-mass star-forming regions, ranging from 2 $\times$ 10$^9$ down to 3 $\times$ 10$^7$ \citep{Godard2010,Gerner2014}, depending primarily on the local volume density of H$_2$ and the cosmic ray ionization rate \citep{Leemker2021}. Possible differences in the HCO$^+$ abundances would influence the derived outflow properties by no more than about an order of magnitude. 
Based on all the above factors, we expect the uncertainty on the final outflow properties to be of the order of a few. 
As the scatter observed in the outflow -- source properties correlations is typically larger than an order of magnitude, the above uncertainties are not expected to significantly alter the subsequent discussion. 

To better put our results into context, we carry out a detailed comparison of the outflow properties calculated here, in the light of the established correlations between outflow and source properties, with an extended literature sample in Section \ref{sec:correlations}.   

\section{Discussion}
\label{sec:discussion}
\subsection{Correlation of outflow and source properties}
\label{sec:correlations}

\begin{figure*}
\sidecaption
      \includegraphics[width=0.29\linewidth]{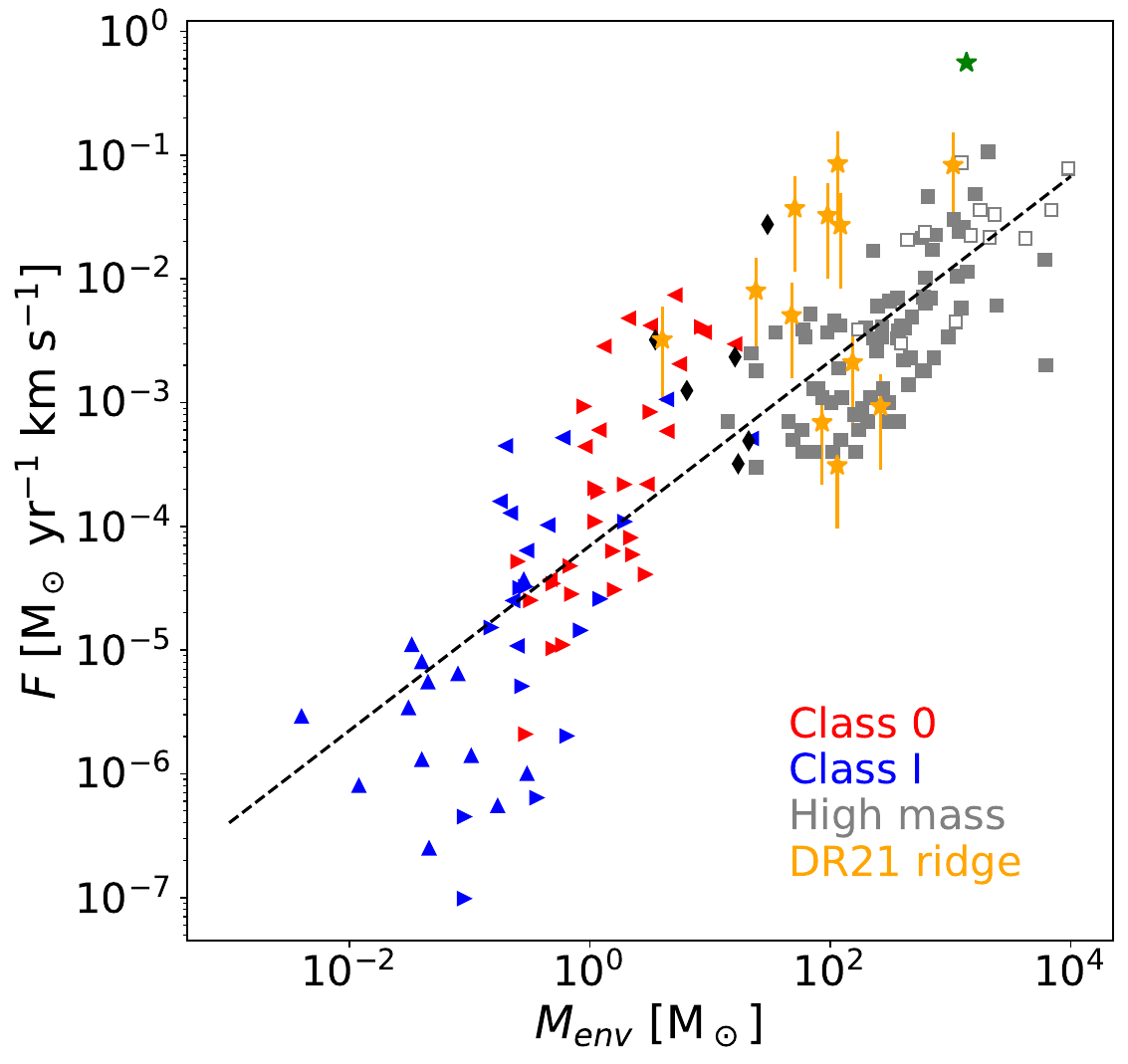}
    \includegraphics[width=0.29\linewidth]{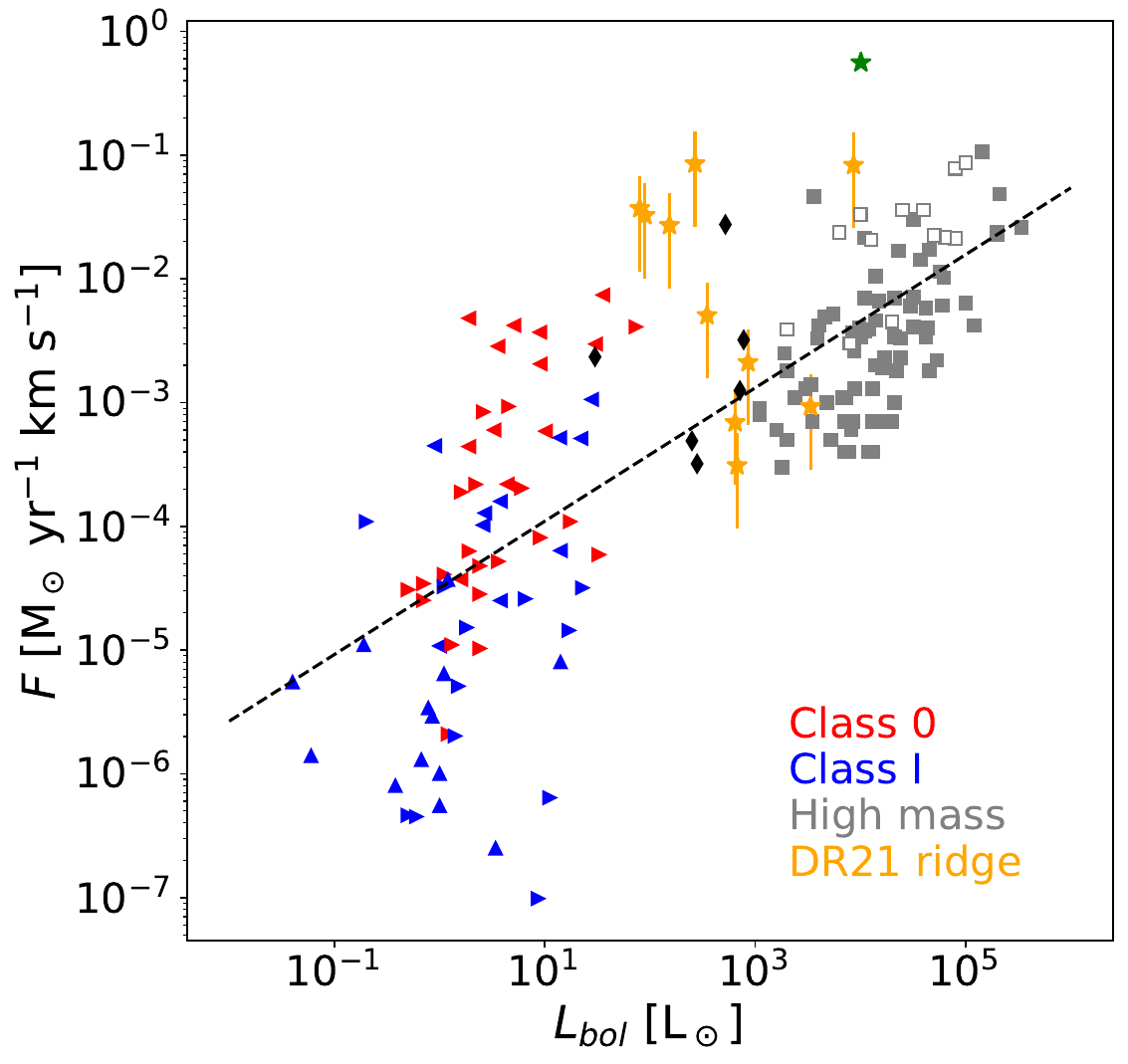}
    \caption{Outflow force over the envelope mass (left) and the bolometric luminosity (right) of the driving protostellar sources. Right facing triangles represent low-mass sources from \citet{Mottram2017}, left facing triangles mark sources taken from \citet{Yildiz2015} and upwards are from \citet{vdm13} while in blue are the Class I sources and in red the Class 0. Black diamonds mark intermediate-mass sources \citep{vk16}, filled gray squares mark high-mass sources from \citet{Maud2015}, clear gray squares mark high-mass sources from \citet{Beuther2002}, and orange stars mark the sources in the DR21 ridge (this work). The green star represents the DR21 Main outflow from \citet{Skretas23} and the dashed, black line shows the best fit to the outflow mass - envelope mass correlation for all sources. The errorbars mark the possible variation of the outflow mass estimate due to the inclination correction factor. }
    \label{fig:fout_corr}
\end{figure*}

The presence of correlations connecting the different energetic properties of protostellar outflows ($M_\text{out}$, $E_\text{kin}$, $P_\text{out}$, and the associated rates) with the properties of their driving sources ($M_\text{env}$ and $L_\text{bol}$) has been well established over the years for both the low- \citep[e.g.,][]{Bontemps1996} and high-mass regimes \citep[e.g.,][]{Beuther2002}. These correlations allow us to better contextualize the outflow properties measured for the outflows in the DR21 ridge by comparing them to previous measurements from the literature.

Figure \ref{fig:fout_corr} shows the correlation between the outflow force ($F_\text{out}$) with both the envelope mass ($M_\text{env}$) and bolometric luminosity ($L_\text{bol}$) for the outflows of the DR21 ridge, along with an extended sample of sources from the literature  \citep{Beuther2002,vk16,vdm13,Yildiz2015,Maud2015,Mottram2017}.
The correlations for the remaining outflow properties are shown in Figs. \href{https://doi.org/10.5281/zenodo.21804190}{E.1 - E.5}. 

For the sources in DR21, the values for $M_\text{env}$ and $L_\text{bol}$ are taken from \cite{Cao19}. N37 and N54 have no counterparts in \cite{Cao19}, therefore $M_\text{env}$ values for these sources are taken from \cite{Mot07}. As \cite{Mot07} does not provide $L_\text{bol}$ measurements, N37 and N54 are not included in the associated correlations. In addition, we note that values in \cite{Cao19} are estimated using SED fits based only on FIR observations, as a result, $L_\text{bol}$ values are likely underestimated.  

The sources of the DR21 ridge cover a range of $\sim2$ orders of magnitude in both the $M_\text{env}$ and $L_\text{bol}$ space, with $M_\text{env} \approx 10 -10^3$ M$_\odot$ and $L_\text{bol} \approx 10^2-10^4$ L$_\odot$. This places the DR21 ridge sources in the poorly populated, intermediate to high-mass regime.   
Within that range of source parameters, the outflows of the DR21 ridge display significant variation in all of their outflow parameter values, often spanning several orders of magnitude. Such scatter is similar to that observed in the low-mass sources, but is not seen in the high-mass sample. Given that the DR21 ridge sources are closer to the high-mass regime, this indicates that the large scatter of the correlation seen in the low-mass regime is likely to actually extend into the high-mass regime. The lack of scatter seen so far in the high-mass samples can potentially be attributed to their, on average, larger distances. More precisely, due to these larger distances, the available spatial resolution, especially of single-dish telescopes, is severely restricted. Disentangling the outflow emission, and accurately identifying the outflow lobes becomes, therefore, significantly more challenging. In addition, the higher distances would lead to a drop in observed fluxes, potentially causing emission from weaker outflows to drop below sensitivity limits. As a result, it is possible that in past surveys of high-mass protostellar outflows, the measured properties are biased, covering only the upper limit of high-mass outflows that were easier to detect at the time. Unbiased observations of high-mass star-forming regions at high enough spatial resolution and sensitivity will potentially reveal whether the lack of scatter in the high-mass regime is real or rather due to observational biases.

Protostellar outflows are closely connected to the accretion process, and by extension to the whole of star formation. Therefore, outflow properties can serve as an easily accessible way to investigate the impact of environmental factors onto the star formation process itself. The DR21 ridge presents a unique case, since it represents the most extreme part of an already intense high-mass star-forming region, Cygnus-X. The close agreement of outflow properties of the DR21 sources with those of the literature sample seen in Fig. \ref{fig:fout_corr} (also in \href{https://doi.org/10.5281/zenodo.21804190}{G.1 - G.5}) suggests that even in regions with exceptional density of star formation activity, such as DR21, the impact of the environment on the outflows and by extension the accretion process is minimal. A similar argument was made by \cite{Beuther2025}, based on studies of multiplicity and the initial mass function (IMF) in different star-forming regions. 
For a more concrete comparison between the DR21 sample and the literature sources, we performed a statistical comparison between the two i.e., the \lq Energy-distance' test \citep{Rizzo2016} in order to identify whether the DR21 sample is consistent with being drawn from the literature sample of outflows. For this test, first the metric $D^2(F,G)$ is calculated as: 
\begin{equation}
    D^2(F,G) = 2E\Vert(X-Y)\Vert - E\Vert(X-X')\Vert - E\Vert(Y-Y')\Vert,
\end{equation}
where $F$ and $G$ are two sets of two-dimensional data points (i.e., the different outflow - source property coordinate pairs for the DR21 sources and the literature sample), $X$ and $Y$ represent elements of $F$ and $G$ respectively (i.e., specific sources in each sample), and $E$ stands for the expected value. 
This value is then compared to the distribution of $D^2$-values estimated by randomly distributing all points from both sets, into new sets with the same sizes as the original and re-calculating the metric. A $p$-value can then be estimated, representing the probability of $D^2$-values greater than the one corresponding to the initial sets being measured. 

The results of the \lq Energy-distance' test for all outflow parameters collected from the literature sample and calculated for the DR21 sources are presented in Tables \ref{table:Edist_menv} and \ref{table:Edist_lbol}.
We find $p$-values $> 0.05$ indicating that there is no statistically significant difference between the DR21 sources and those drawn from the literature. We note, however, that the sample size of the DR21 sources is comparatively small, when contrasted to the sample of literature sources, and therefore small differences in the sample distributions might not be as noticeable in the statistical comparison. Despite that, the good agreement of the DR21 sources with the broad literature sample, with only exception the explosive outflow candidate of DR21 Main, suggests that protostellar outflows, and by extension the accretion and star-formation process, remain largely unaffected even when taking place in extreme and clustered environments, such as the DR21 ridge. 

The logarithms of outflow properties have been found to scale linearly with those of the source properties \citep[e.g.][]{Beuther2002,vdm13,Skretas22}. With the addition of the DR21 ridge sources, we carry out new linear fits for all outflow properties correlations with both $M_\text{env}$ and $L_\text{bol}$ providing updated values for the scaling relations. The resulting best-fit parameters are shown in Tables \href{https://doi.org/10.5281/zenodo.21804190}{F.1-F.6}. The fits are carried out separately for low-, intermediate- and high-mass sources, but also for the combined sample. Among the individual samples, the tighter correlations (higher Pearson's $r$ coefficients) are found for the low-mass sample. In contrast, the intermediate-mass sample displays the lowest $r$ coefficients, likely due to the limited sample size. Notably, in some cases the intermediate-mass sources display negative $r$ values, suggesting an anti-correlation with the outflow properties. In Fig. \ref{fig:fout_corr}, some of the DR21 sources can be seen to better match the top end of the distribution of low-mass sources, while others align better with the lower end of the high-mass objects distribution. This reinforces the previously observed discrepancy between high- and low- mass sources in terms of the correlations between outflow and driving source properties \citep[e.g.,][]{Skretas22}. We stress though that the sample of intermediate mass sources still remain relatively limited in comparison to both the high- and low-mass samples and further observations of such objects are required to confirm whether the observed split is real, or the result of missing observations.
When the entire sample of protostellar outflows is considered together, a tight, positive linear correlation is always found with Pearson's coefficients $\ge0.6$. The existence of this tight correlation across such an extended range of source properties (five orders of magnitude in $M_\text{env}$ and 6 in $L_\text{bol}$), supports the existence of a common formation mechanism for both low- and high-mass stars.        

Summarizing the findings of this section, the comparison of the outflow and source properties correlations reveals that the outflows of the DR21 ridge occupy mainly the intermediate to high-mass range. In addition, they display a significant scatter, similar to that of low-mass sources, suggesting that this scatter extends into the high-mass regime, where it was not clearly seen previously. Finally, the outflow properties of the DR21 sample appear statistically similar to those of the extended sample, suggesting that even the extreme environment of DR21 has no noticeable impact on the star formation process.    

\begin{table}[]
\caption{Energy distances ($E_\text{dist}$) and corresponding p-values for energy distance tests of the distributions of outflow parameters with respect to the source envelope mass.} 
\label{table:Edist_menv}
\centering
\small
\begin{tabular}{l c c} 
\hline\hline 
 & $E_\text{dist}$& $p-$value \\ 
\hline
Outflow Mass & 77.06 & 0.29 \\
Mass rate & 71.99 & 0.31 \\ 
Momentum & 122.36 & 0.17 \\
Force & 71.99 & 0.31 \\ 
Kinetic energy & 4.10$\times 10^{45}$ & 0.24 \\ 
Kinetic luminosity & 71& 0.3160 \\ 
 \hline
\end{tabular}
\tablefoot{In the cases where p-value > 0.05, the samples cannot be differentiated statistically.}

\end{table}

\begin{table}[]
\caption{Energy distances ($E_\text{dist}$) and corresponding p-values for energy distance tests of the distributions of outflow parameters with respect to the source bolometric luminosity.} 
\label{table:Edist_lbol}
\centering
\small
\begin{tabular}{l c c} 
\hline\hline 
 & $E_\text{dist}$& $p-$value \\ 
\hline
Outflow Mass & 6550.41621 & 0.09 \\
Mass rate & 6550.39 & 0.09 \\ 
Momentum & 6551.94 & 0.09 \\
Force & 6550.39 & 0.09 \\ 
Kinetic energy & 3.72$\times 10^{45}$ & 0.38 \\ 
Kinetic luminosity & 6548.04 & 0.10 \\ 
 \hline
\end{tabular}
\tablefoot{ In the cases where p-value > 0.05, the samples cannot be differentiated statistically.}

\end{table}

\subsection{Distribution of outflow sources along the ridge}

\begin{figure*}
    \centering
    \includegraphics[width=0.3\linewidth]{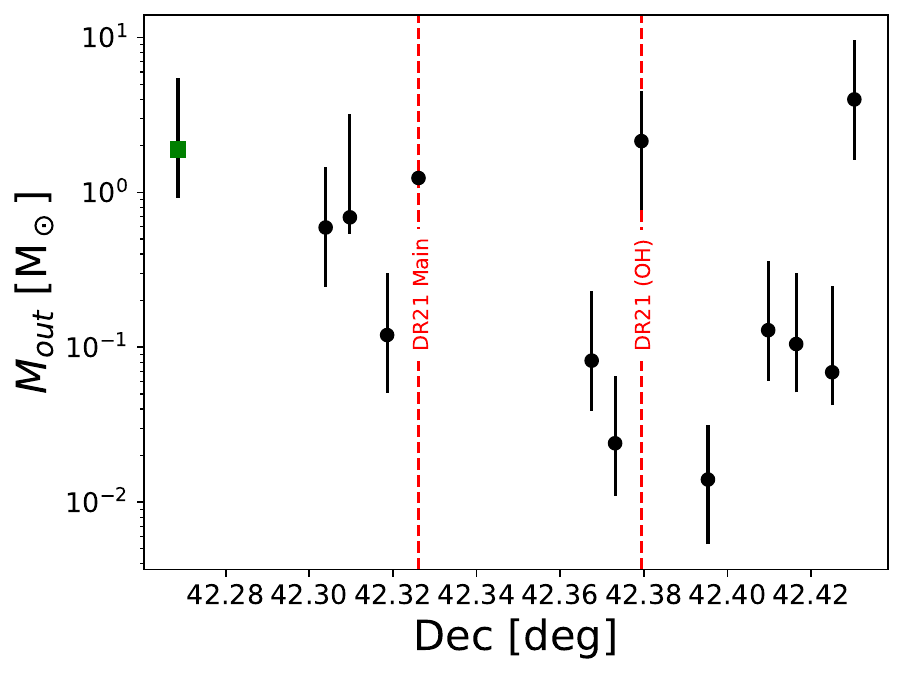}
    \includegraphics[width=0.3\linewidth]{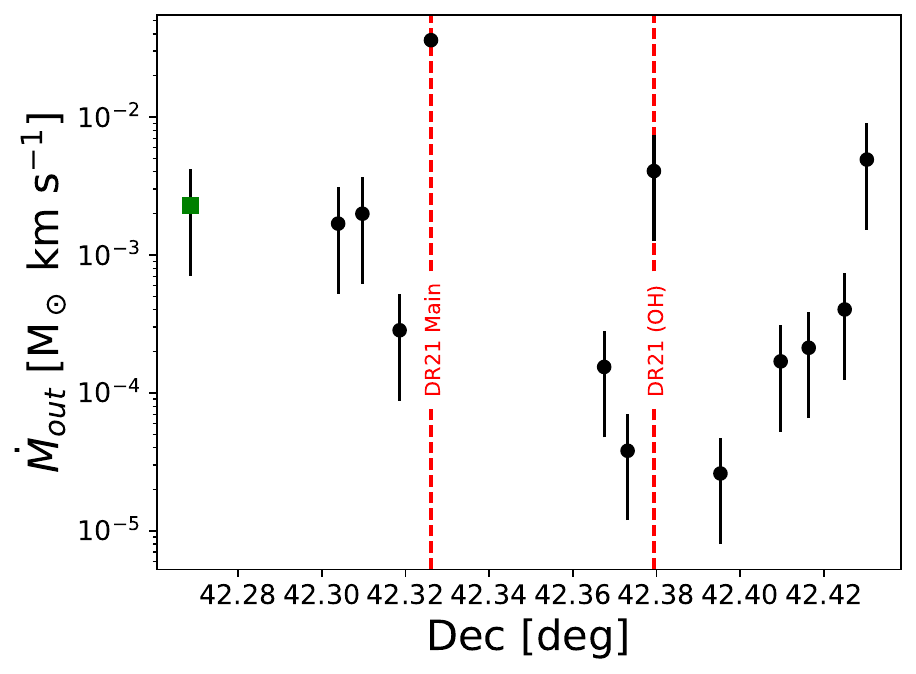}
    \includegraphics[width=0.3\linewidth]{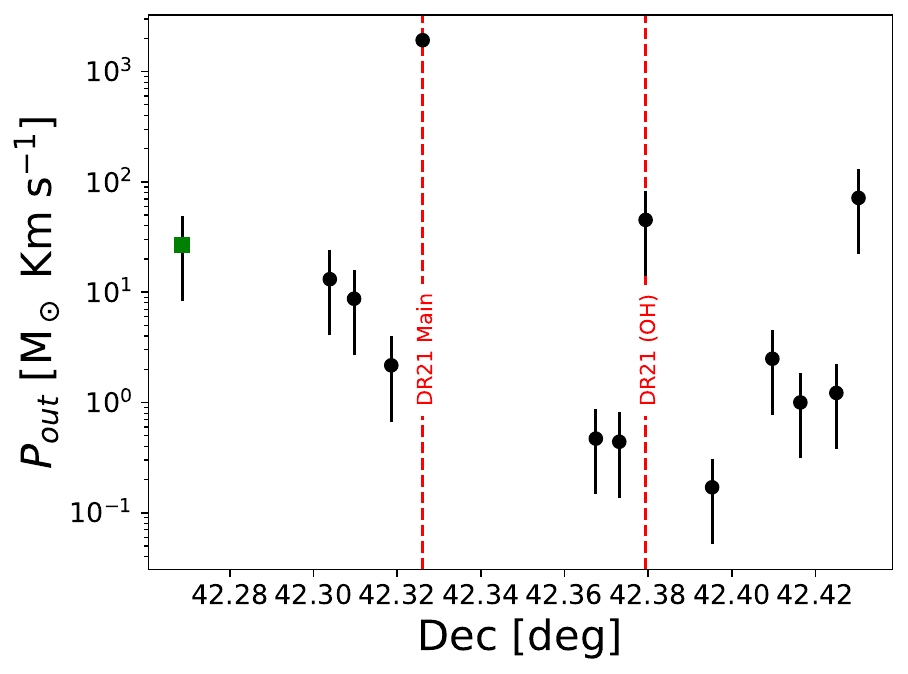}
    \includegraphics[width=0.3\linewidth]{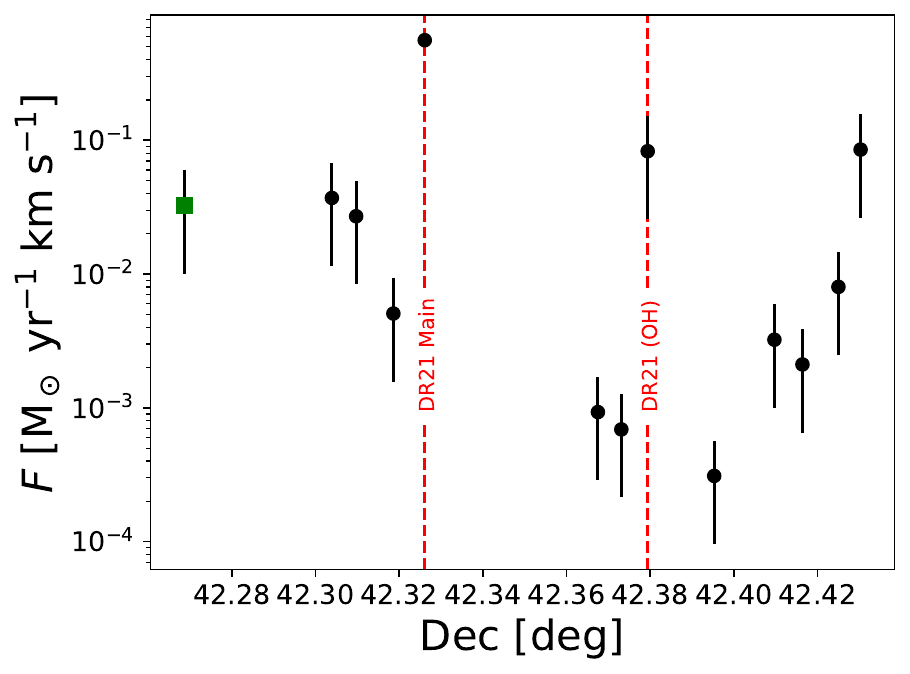}
    \includegraphics[width=0.3\linewidth]{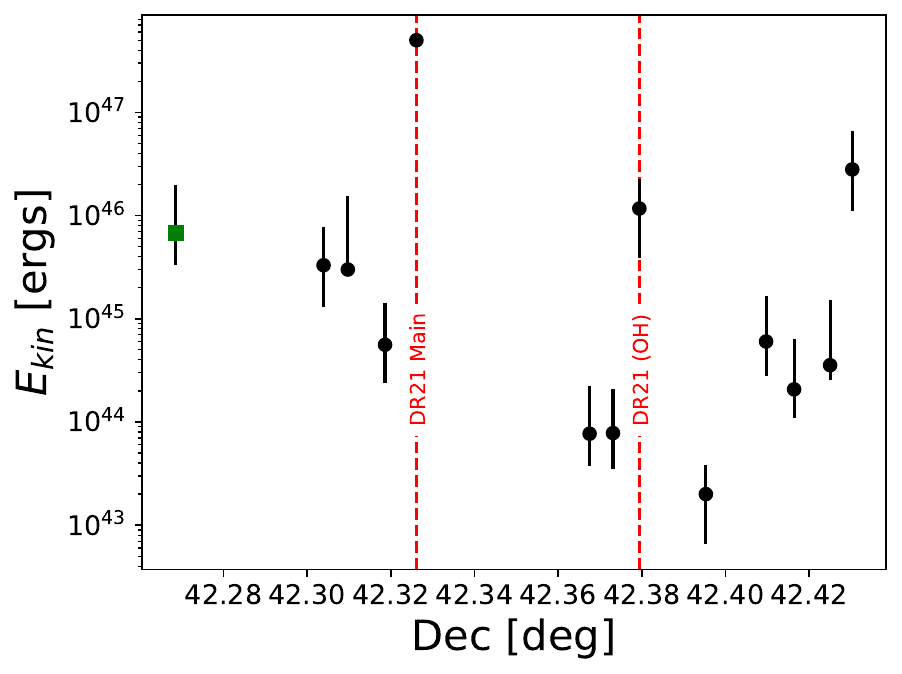}
    \includegraphics[width=0.3\linewidth]{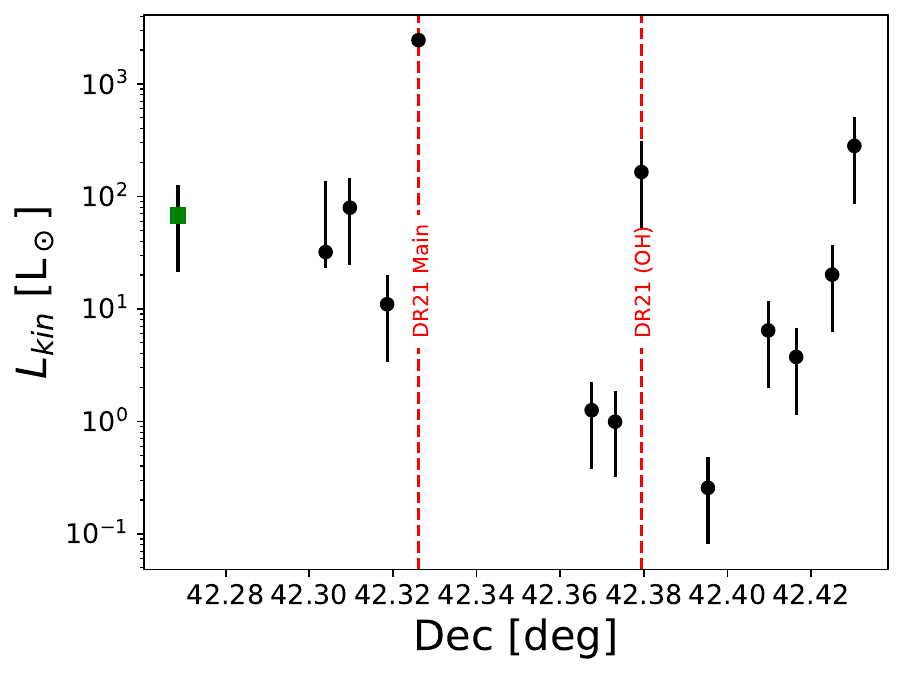}
    \caption{Outflow properties over source declination for all outflow sources in the sample. From top to bottom the outflow properties shown are: outflow mass (left) and outflow mass rate (right), outflow momentum (left) and outflow force (right), and outflow kinetic energy (left) and luminosity (right). Dashed lines mark the location of DR21 Main and DR21 (OH). The green square corresponds to the outflow of N56, which is located in a separate structure from the DR21 ridge. The errorbars show the impact of the inclination correction on the properties. 
    }
    \label{fig:properties}
\end{figure*}

The impact of the general environment (e.g. strong UV fields from nearby OB stars, clustered versus isolated formation, physical conditions of the star-forming molecular cloud) onto the process of star formation is still a matter of debate \citep[e.g.,][]{Beuther2025}. In this section, we discuss our results regarding the outflows of sources along the DR21 ridge in the context of the large-scale properties of the filament in an effort to constrain whether they are impacted in any way from the broader environment. 

\begin{figure}
    \centering
    \includegraphics[width= 0.7\linewidth]{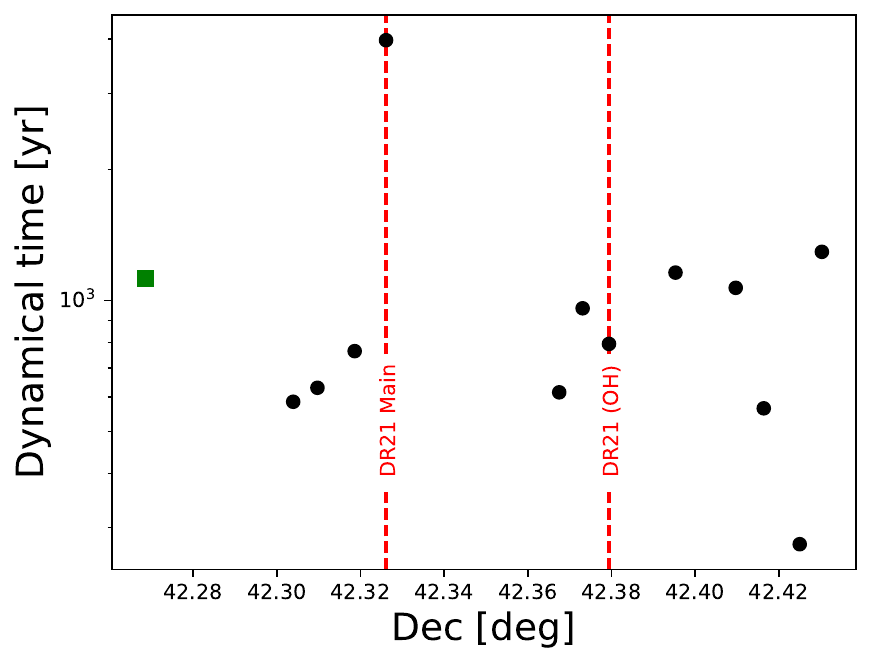}
    \caption{Dynamical time over source declination for all outflows sources in our sample. The dynamical time shown here is the average between the dynamical times estimated for the two outflow lobes. Dashed lines mark the location of DR21 Main and DR21 (OH). The green square corresponds to the outflow of N56, which is located in a separate structure from the DR21 ridge.}
    \label{fig:tdyn}
\end{figure}

Previous $\textit{Herschel}$ observations of the DR21 ridge revealed the presence of a gradient in the evolutionary stage of the various dense clumps located along the filament. \cite{Hen12} found that both the 70 $\mu$m luminosity and the dust temperature of the dense clumps along the DR21 ridge decrease moving from the South to the North end of the filament. We investigate here whether this evolutionary trend along the DR21 ridge has a noticeable impact on the protostellar outflows.
Figure \ref{fig:properties} shows the distribution of outflow properties along the DR21 ridge revealing a lack of any obvious trends. Three sources appear to stand apart from the majority in terms of their outflow properties, especially the different rates (i.e., $\dot{M}_\text{out}$, $F$, and $L_\text{kin}$), DR21 Main-N46, DR21 (OH)-N44 and N53. These three sources are relatively equally spaced along the DR21 ridge, with DR21 Main on the south end, DR21 (OH) in the middle, and N53 towards the northern edge. When not taking these exceptional cases into account, a decrease in outflow properties can be noted between Dec $\sim 42.3$$\degr$ and $\sim42.4$$\degr$; however, outflow properties seem to increase again as we move further North. The strong correlation of the outflow properties with those of the driving source (see Sec. \ref{sec:correlations}) is potentially overshadowing any trend arising  from the evolutionary stage of the clumps along the ridge.
Looking at the dynamical times of the outflows along the ridge, as a more direct estimate of the outflow age (Fig. \ref{fig:tdyn}), we again find no sign that the outflows along the ridge follow the reported evolutionary trend. Interestingly, sources in the south end of the ridge, associated with relatively higher outflow properties have relatively low dynamical times, while on the other hand, DR21 Main and N53 the two sources with consistently highest outflow properties also have the highest dynamical ages. There does not seem to be a direct connection between the dynamical time and the outflow properties. We note here though that the dynamical time of an outflow is a highly inaccurate measure of its actual age, due to the impact the inclination angle has on its calculation. Overall, we find no indication that the North to South evolutionary trend of the DR21 ridge has any impact on the protostellar outflows themselves. 

\begin{figure}
    \centering
    \includegraphics[width=0.9\linewidth]{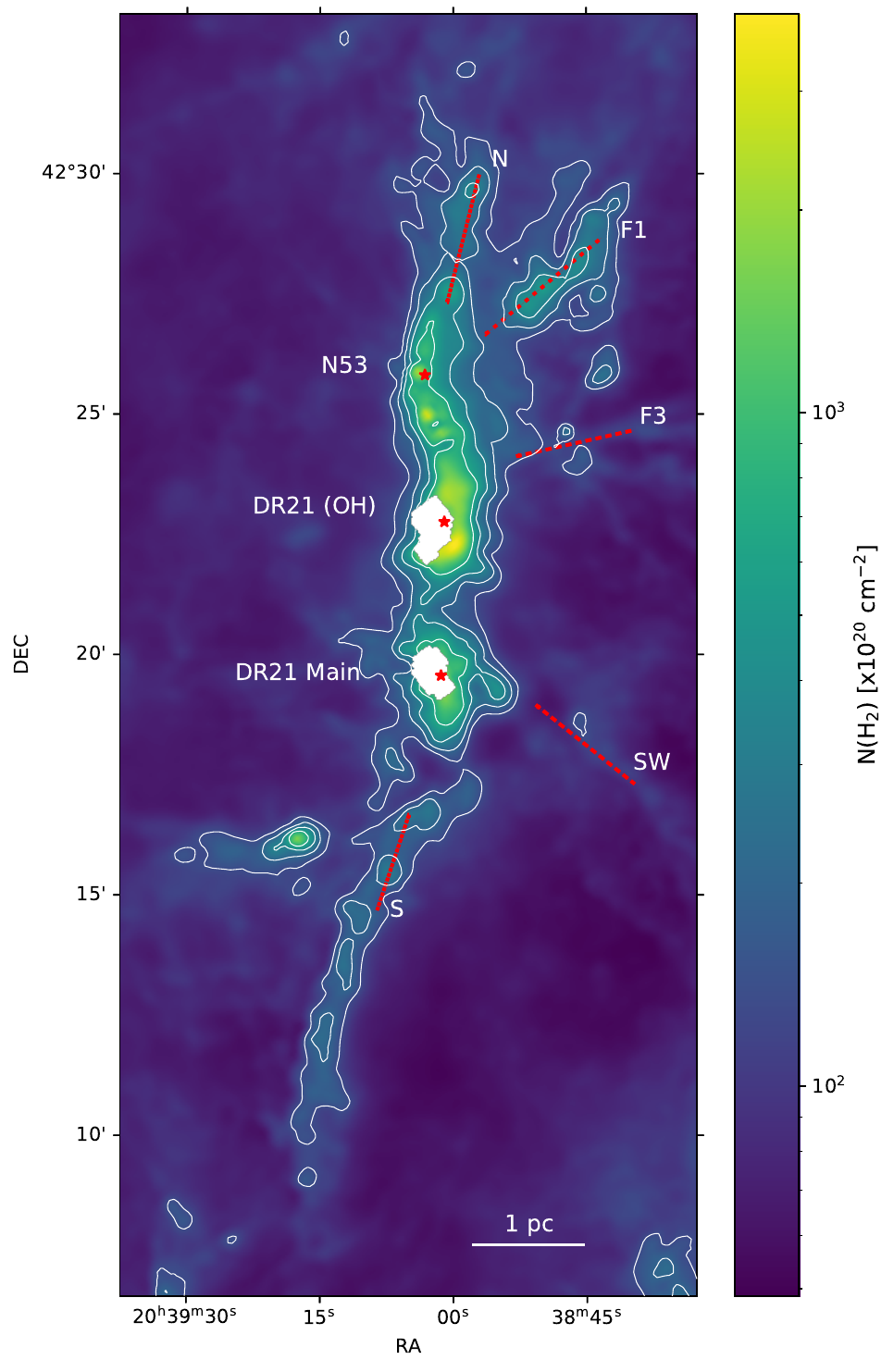}
    \caption{H$_2$ column density map of the entire DR21 ridge from \cite{Marsh17}. Contour levels correspond to 2,3,5, and 10 $\sigma_\text{rms}$ of the column density. Red stars mark the location of the most prominent outflow sources DR21 Main - N46, DR21 (OH)- N44, and N53. Red dotted lines mark the location of the accreting filaments, as identified by \cite{Hen12}. The saturated regions around DR21 Main and DR21 (OH) have been masked.}
    \label{fig:coldens}
\end{figure}

Another interesting characteristic of the DR21 ridge is that it appears to still be accreting mass, through multiple filaments connecting to it \citep{Schneider10,Hen12,Pillai2026}. The intersection of such filaments has often been proposed as the location of high-mass star formation \citep[e.g.,][]{Balsara01,Banerjee06,Smith11,Mot18,Kumar2020,Hacar25}
Indeed, in the case of DR21, it has been shown that the F3 filament connects to DR21 (OH) \citep{Csengeri11}, while the southern filaments S and SW are associated with DR21 Main \citep{Hen12}. This can also be seen in Fig. \ref{fig:coldens} where the location of the filaments reported in \cite{Hen12} is shown over the column density map of the DR21 ridge from \cite{Marsh17}. In addition, Fig. \ref{fig:coldens}  shows that the massive clump N53 is located close to the position where the two northern filaments F1 and N would intersect.  
While a more detailed analysis of the filaments is required to firmly confirm the connection between the filaments and the clumps, based on the spatial distribution of the clumps and previous studies \citep{Csengeri11,Hen12} all three most outstanding outflow sources along the ridge appear closely associated with filaments accreting material.

The strongly skewed morphology of the DR21 ridge \citep[e.g.,][]{Bonne2023} presents a great opportunity to investigate whether the large-scale morphology of a star forming filament impacts the direction of rotation of the dense cores forming within it.
As protostellar outflows are launched from the accretion disk surrounding the protostar, their orientation is in principle determined by that of the accretion disk. 
While the orientation of outflows can potentially shift during the evolution of a protostar \citep[e.g.,][]{Offner2016}, at the earlier stages it is likely to depend on the original rotation of the collapsing core and subsequent accretion disk. The direction of protostellar outflows arising from the dense cores along the DR21 ridge could shed some light on whether core rotation is influenced by the large-scale cloud morphology and kinematics or is instead dictated by more local dynamics. 
 
\begin{figure}
    \centering
    \includegraphics[width=0.8\linewidth]{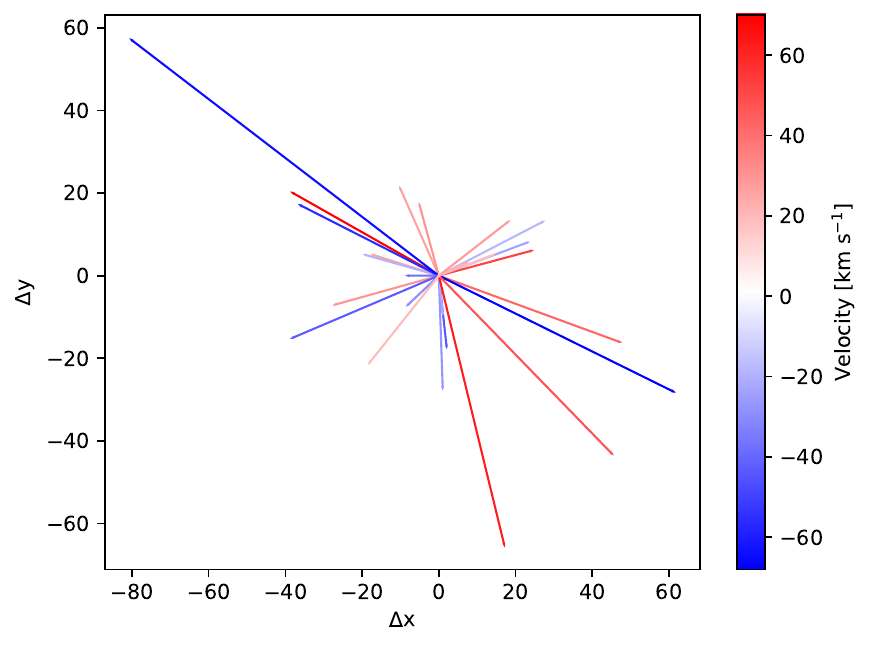}
    \caption{Orientation and extent of all outflow lobes in the DR21 ridge with the exception of those of the DR21 Main outflow. The colorscale of the lines corresponds to the maximum/minimum velocity measured for red- and blue-shifted lobes respectively.}
    \label{fig:outflow_direction}
\end{figure}

Figure \ref{fig:outflow_direction} shows that outflows along the DR21 ridge display no preferential direction, but appear more randomly distributed, suggesting the lack of a preferred core rotation direction.
An argument can be made that some of the shorter outflows could actually be closer to the line-of-sight direction, instead of north-south and thus support a preference of outflows to extend perpendicular to the filament, instead of along it. In that case, we would expect such outflows to display higher velocities, since the projection effect would be less significant. However, shorter outflows show lower velocities,  indicating that this is not the case (Fig.~\ref{fig:outflow_direction}). While our sample is limited to a single cloud, our results suggest that there is no strong connection between the orientation of the large-scale filamentary structure and the rotation of the dense cores forming within it, similar to previous results from low-mass protostars in the Orion and Perseus molecular clouds by \citet{Davis2009,Stephens2017}.       

In brief, the comparison of outflow properties between the sources along the DR21 ridge showed that the established evolutionary trend along the filament has no noticeable impact on the current protostellar outflows.
Finally, despite the strongly skewed large-scale morphology of the DR21 ridge, there appears to be no preferential rotation direction passed down to the core scales, as the outflow directions appear spread uniformly towards all directions.

\subsection{Comparison of CO and HCO$^+$ outflows}
\label{sec:covshco+}

\begin{figure*} 
    \includegraphics[width=0.33\linewidth]{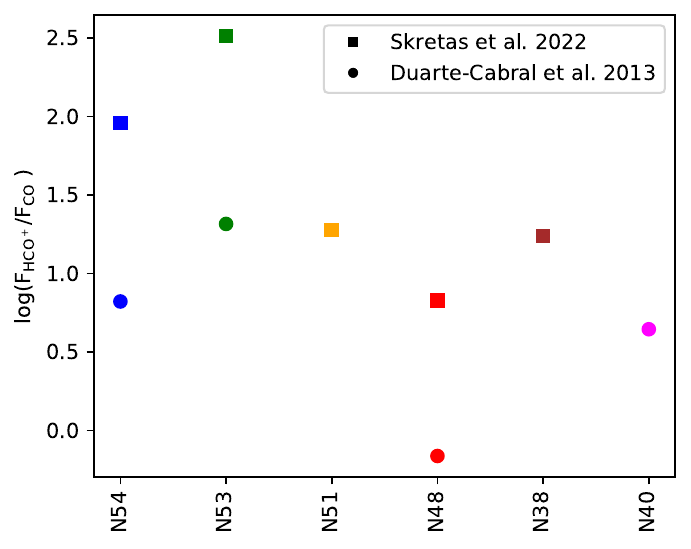}
    \includegraphics[width=0.33\linewidth]{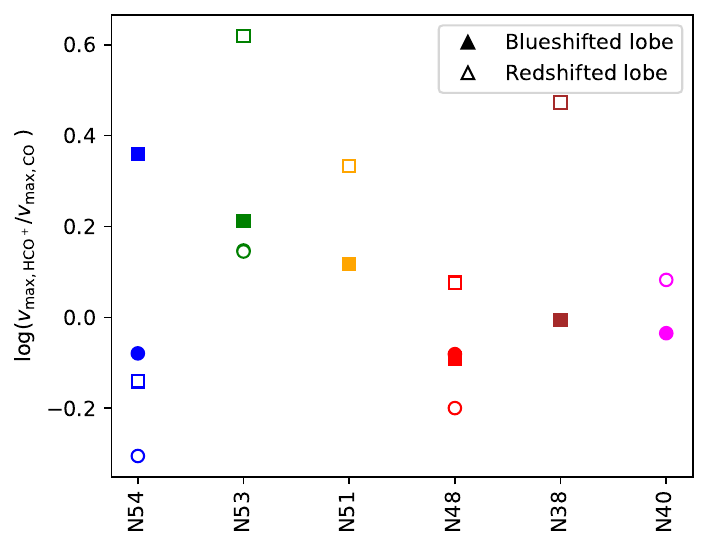}
    \includegraphics[width=0.33\linewidth]{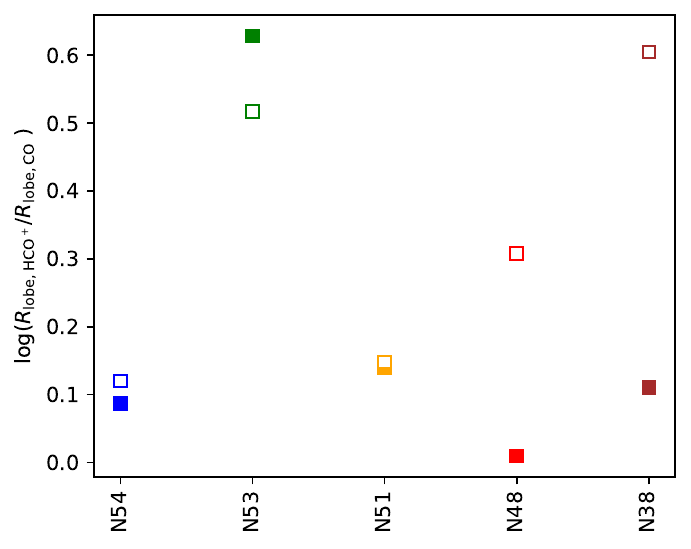}
    \caption{Comparison of different outflow properties measured using the CASCADE HCO$^+$ observations (this work) and CO $J=3-2$ observations taken with the SMA \citep{Skretas22}.
    [Top]: Comparison between the outflow forces ($F$). [Middle]: Maximum velocity ($v_\mathrm{max}$) in each lobe. [Bottom]: Radius of outflow lobes (R$_\mathrm{lobe}$). Square points, represent comparisons with outflow values from \citet{Skretas22}, while circles show comparisons with \citet{duarte13}. In the second and third panels, filled and clear markers correspond to the values of the blue- and red-shifted outflow lobes respectively. }
    \label{fig:COvsHCO}
\end{figure*} 

The outflow properties in this work are estimated using HCO$^+$ $J=1-0$, whereas all previous detections of outflows in DR21 ridge used low-$J$ transitions of $^{12}$CO (Section \ref{sec:correlations}).
Due to the relatively high critical density of HCO$^+$ with respect to $^{12}$CO \citep[e.g., $n_\text{crit} \sim 1.5 \times 10^{5}$ cm$^{-3}$ for HCO$^+$ $J=1-0$, and $\sim 2.2 \times 10^3$ cm$^{-3}$ for CO $J=1-0$][]{Yazidi2014, Bolatto2013}, it is possible that these two molecular species do not trace the same outflow components. Indeed, it has been suggested that HCO$^+$ might be more closely associated with the outflow cavity walls than the bulk of the molecular outflows \citep[e.g.,][]{Rawlings2004,Arce2006}. To validate the comparisons of Sec. \ref{sec:correlations}, it is therefore crucial to investigate the impact that the use of different molecular tracers has on the outflow properties. 
We compared the outflow forces obtained using HCO$^+$ with previous measurements from the literature, that used observations of CO $J=2-1$ \citep{duarte13} and $J=3-2$ \citep{Skretas22}. 
Figure \ref{fig:COvsHCO} shows the ratio between the outflow forces, maximum outflow velocities, and outflow extents measured in this work and those from \citet{duarte13} and \citet{Skretas22} for sources with both CO and HCO$^+$ observations (N54, N53, N48, and N40 for \citet{duarte13} and N54, N53, N51, N48, and N38 for \citet{Skretas22}). As mentioned also in Section \ref{sec:outflow_props}, the values for the outflow forces calculated here are found to be $\sim$1 - 2 orders of magnitude higher than those from \cite{Skretas22} and up to an order of magnitude higher than those found in \citet{duarte13}. 
The lack of short spacing observations in the PILS-Cygnus survey, leading to significant filtering of the extended outflow emission could partially explain the very large difference seen in that case. This is also reflected in the $v_\text{max}$ and $R_\text{lobe}$ values measured in the two studies, with HCO$^+$ measured values being usually higher by a factor of a few. 
At the same time though, our results appear also typically higher than those from \citet{duarte13}, where short spacings information is included, suggesting that our HCO$^+$ $J=1-0$ observations are potentially probing slightly different parts of the outflow compared to the CO studies.

In fact, comparisons of outflow properties using CO and  HCO$^+$ in the literature provided contradictory results. For example, \cite{WS14} found outflow properties of low-mass protostars of about an order of magnitude lower using HCO$^+$ $J=4-3$ than those from a previous study using $^{12}$CO $J= 3-2$ \citep{Curtis10}. In contrast, \cite{Liu21} reported that outflow properties measured using HCO$^+$ 1-0 are higher by approximately an order of magnitude than those measured with $^{12}$CO for a sample of high-mass sources. The use of different transitions of the two molecules can have a significant impact on the resulting outflow properties, as the critical densities of each transition, even for the same molecule, are different and therefore can probe different parts of the outflows. As a result, the reported changes in outflow properties can also be at least partially attributed to these effects.
It becomes apparent that, additional comparisons between the two outflow tracers, more consistent in terms of observations, are required in order to accurately constrain the relation between them. To that end, the NASCENT-stars project \citep{Csengeri2025} can potentially allow for a significantly more accurate and thorough comparison of properties between CO and HCO$^+$ outflows, by providing CO observations toward the regions covered by CASCADE.
Despite this uncertainty, due to the large scatter inherent in the correlations of outflow and source properties, the conclusions of Sec. \ref{sec:correlations} would remain largely unaffected even if the outflow properties were to shift by an order of magnitude in either direction.

\section{Summary and conclusions}
\label{sec:conclusions}
In this work, we presented HCO$^+$ $J=1-0$, H$^{13}$CO$^+$ $J= 1-0$, and SiO $J=2-1$ observations covering the entire DR21 ridge. The observations were taken as part of the CASCADE project, using both the NOEMA interferometer and the IRAM 30m telescope. We took advantage of the extended coverage and great spatial resolution of the CASCADE observations to investigate the HCO$^+$ emission associated with all dense molecular cores along the filament for signs of molecular outflows. Subsequently, we compared the outflow activity inside the ridge both with the large-scale properties of the filament as well as with extended outflow samples from the literature to assess the impact of environment onto the outflow properties and by extension the star formation process. Our conclusions are as follows: 

\begin{itemize}
    \item Based on our HCO$^+$ emission and moment 0 maps, we report the detection of outflows in 14 out of 34 dense molecular cores in the DR21 ridge. The outflows appear bipolar in all but one case.
    \item We calculated outflow properties (mass, momentum, kinetic energy and the related rates) for all detected outflows using the Separation method \citep{vdm13}. We find the resulting outflow properties of the sources in the DR21 ridge not to be statistically different compared to those of an extended literature sample, suggesting that the extreme and clustered environment of the DR21 ridge has no significant impact on the outflow properties. The close connection of outflow activity with accretion suggests that the star formation process is not strongly affected by the broader environment. 
    \item The DR21 sample extends the currently available sample in the intermediate-mass regime, assisting in bridging the apparent gap between high- and low-mass sources in the outflow - source properties correlations. 
    \item We find no signs of a correlation between the outflow properties and the declination of the source, suggesting that the reported evolutionary stage gradient along the DR21 ridge \citep{Hen12} does not affect significantly the ongoing star formation. 
    \item We see no preferential direction of the bipolar outflows along the DR21 ridge, suggesting that the large-scale morphology does not strongly impact the core scales that dictate the outflow launch direction. 
\end{itemize}

Overall, our results indicate that the star formation process, even in extreme and clustered environments such as the DR21 ridge, remains largely unaffected by the broader environment and is rather dictated by local processes. In addition, our results support the presence of a single, common correlation between outflow and source properties across all $M_\text{env}$ and $L_\text{bol}$ ranges. Additional observations of protostellar outflows, more consistent in term of tracers, resolution, and sensitivity are required to solidify this result. The existence of a common correlation across the entire $M_\text{env}$ regime would strongly suggest a single formation mechanism for stars of all masses.    

\section*{Data availability}
Appendices D-G are available online on \href{ https://doi.org/10.5281/zenodo.21804190}{Zenodo}

\begin{acknowledgements}
The authors would like to thank the anonymous referee for his suggestions and constructive comments that improved the overall quality of the manuscript.
The authors are also grateful to the staff at the NOEMA and Pico Veleta observatories for their support of these observations. We thank in particular P. Chaudet, operator at the NOEMA observatory, for his motivation and dedication in developing and testing the advanced mosaic observing procedures employed in this project. This work is based on observations carried out under project number L19MA with the IRAM NOEMA Interferometer and [145-19] with the 30 m telescope. IRAM is supported by INSU/CNRS (France), MPG (Germany) and IGN (Spain). A.~K. acknowledges support from the Polish National Science Center SONATA BIS grant No. 2024/54/E/ST9/00314. 
N.S. acknowledges support from the CRC 1601 (SFB 1601 sub-project B2) funded by the DFG – 500700252 and DYNAVERSE.
D.~S. was funded by the Deutsche Forschungsgemeinschaft (DFG, German
Research Foundation) – project number: 550639632. 
\end{acknowledgements}

\bibliographystyle{aa}

\bibliography{biblio.bib}

\appendix

\section{Channel maps of the different velocity componenents of DR21}

\begin{figure*}
    \centering
    \includegraphics[width = 0.9\linewidth]{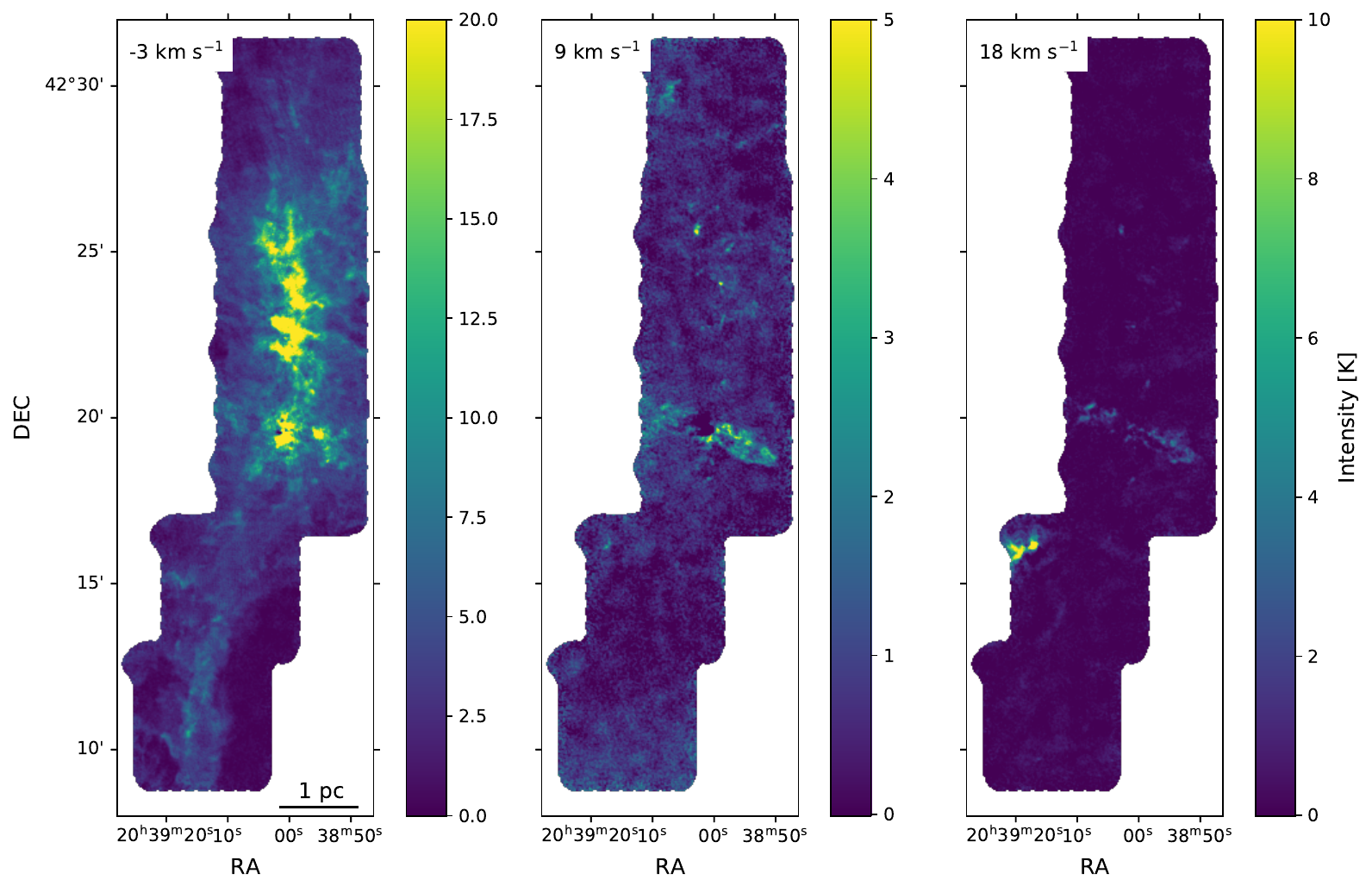}
    \caption{HCO$^+$ channel maps of the 3 different velocity components detected in the DR21 ridge. The main component at -3 km s$^{-1}$ is shown on the left panel, the 9 km s$^{-1}$ component is shown in the middle, while the 18 km s$^{-1}$ component is on the right. Note that the extended emission at declination of $\sim20'$ seen in both the 9 and 18 km s$^{-1}$ maps comes from the outflow of DR21 Main.}
    \label{fig:hco+_components}
\end{figure*}

Figure \ref{fig:hco+_components} displays single channel intensity maps of the HCO$^+$ emission at the velocities of  -3, 9, and 18 km s$^{-1}$, corresponding to the three separate velocity components identified in the DR21 ridge.

\section{Description of outflow sources}
\label{app:outflowsources}
We describe here in more detail the HCO$^+$ outflow sources identified in Section \ref{sec:outflow_maps}. The corresponding integrated intensity maps are shown in Fig. \ref{fig:outflowmaps1}.

\paragraph{N37:}
No clear continuum peak, associated with N37, can be seen in the 3.6 mm emission map. Nevertheless, we report the tentative presence of a bipolar outflow, extending along the E-W direction. The red lobe appears brighter, and extending to higher velocities compared to the blue lobe. Interestingly, the red lobe appears clearly detached from the N37 core. This, combined with the presence of SiO emission at the forward edge of the red lobe, suggests that the HCO$^+$ emission is likely tracing compressed, entrained gas at the front of the outflow. The relative weakness of the blue lobe might arise from the characteristics of the environment surrounding N37. Namely, the blue lobe extends towards the dense material associated with N43, which is likely to rapidly slow down the outflowing material. Another possibility is that higher column densities around N43 lead to the attenuation of the blue-shifted emission.

\paragraph{N38:}
There is a tentative detection of an HCO$^+$ outflow, extending in the E-W direction. The outflow appears to be relatively weak, especially in the case of the red lobe, which is barely detected, and significantly detached from the most likely origin point of the outflow. Even so, the proposed orientation of the outflow, shown in Fig. \ref{fig:outflowmaps1}, closely matches that of the CO outflow reported by \cite{Ching18}, leading us to believe that the HCO$^+$ emission seen is indeed part of a molecular outflow. An additional CO outflow, not seen in HCO$^+$, was reported in \cite{Skretas22}. This second outflow, extends in a North-South direction and appears to match well the SiO outflow previously reported in \cite{Ching18} and \cite{Yang24}.

\paragraph{N40:}
Two possible outflow configurations are marked in the case of N40. The first configuration consists of the stronger detections in both red- and blue-shifted emission (solid arrows, Fig. \ref{fig:outflowmaps1}). An additional outflow configuration, though, extends on the East - West direction, sharing the same blue-shifted lobe as the first outflow candidate (dotted arrows, Fig. \ref{fig:outflowmaps1}). This configuration matches the outflow seen in SiO emission both in this work and also in \cite{duarte14}. However, no HCO$^+$ outflow is seen at the location of the small-scale CO outflow, reported in \cite{duarte13}. In addition, an extended red-shifted and a less extended blue-shifted structure are seen at the north edge of the FWHM of N40. The origin of these structures is unclear, but they do not seem to be associated with continuum emission from the N40 core. As a result, they are not taken into consideration in the analysis of molecular outflows associated with dense cores.

\paragraph{N42:}
The area North of N42 is dominated by emission from the DR21 Main outflow. To the East of N42, the known blue-shifted structure, extending from DR21 Main towards the South can also be seen. Between these two large-scale structures, a bipolar outflow extending mostly in the N-S direction, is detected. This structure could potentially be part of the extended DR21 Main outflow, but given that both the blue and red lobes appear to peak over the area associated with N42, it is most likely that the origin of the emission comes from within N42. To the South of N42, the outflow associated with N50 is also visible. 

\paragraph{N44:}
N44 corresponds to DR21(OH), the second most prominent continuum source in the DR21 ridge, after DR21 Main. Two prominent continuum peaks can be seen within the FWHM of N44, which fragment into multiple peaks at higher resolution observations \citep{zapata12}. The HCO$^+$ emission closely follows that of the CO emission \citep{zapata12}. The presence of multiple continuum peaks, as well as the separation seen predominantly in the red-shifted lobes, supports the scenario of two separate outflows in N44, as was proposed by \cite{zapata12}. At the same time, however, we cannot rule out the scenario that the HCO$^+$ emission is tracing the cavity walls of a large-scale outflow. 
A large-scale methanol outflow, with opposite orientation from that of the CO and HCO$^+$ outflows, has been reported by \cite{araya09}. A significant blue-shifted structure, seen at the eastern edge of N44 could potentially be associated with this outflow, but no corresponding red-shifted emission is detected. For the purposes of this work, only the outflow structure that matches the CO outflows from the literature will be considered.

\paragraph{N45:}
A bipolar outflow, extending in the E-W direction is clearly seen towards the southern half of the FWHM of N45. An additional blue-shifted structure is visible to the SW of N45. At lower velocities this structure appears to be loosely connected to the blue-shifted emission that extends southwards from DR21 Main. The outflow associated with N50 is also visible North of N45. 

\paragraph{N46 \& N47:}
The cores of N46 and N47 are part of the DR21 Main region. The powerful outflow of DR21 Main is clearly seen in HCO$^+$ emission, and has been discussed in detail in \cite{Skretas23}. As a uniform size of 0.5$\times$0.5 pc was used for all the outflow maps presented in this work, the outflow of DR21 Main extends outside the limits of the presented field in both the East and West directions. Given the more central location, the closest association with SiO emission, the stronger and more extended continuum emission, the outflow of DR21 Main is most likely originating from N46. A possible scenario for the formation of the DR21 Main outflow is that it actually consists of multiple different outflows, all aligned in the same direction. In such a scenario, given the complexity of the emission in the region, we cannot rule out the presence of an outflow originating from N47. For the remainder of this work though, it will be assumed that the outflow of DR21 Main is driven by N46. 

\paragraph{N48:}
There is a tentative detection of a blue and red outflow lobes close to the continuum peak of N48. While the HCO$^+$ detection is not very clear, with emission seen mostly at lower velocities, its orientation appears to match that of the CO outflow \citep{duarte13,Ching18,Skretas22}, suggesting that the emission indeed originates from the bipolar outflow. An additional red-shifted structure is seen at the South-West part of the FWHM of N48. This structure is considered as an additional red-shifted outflow in \cite{Ching18}, but it does not appear connected to any of the continuum fragments \citep{Bon10}, and also lacks a blue-shifted counterpart. As a result, we excluded it in further analysis in this work. The location of the infrared source IS-1, a more evolved source detected with IRAC at 3.6 and 4.5 $\mu$m is marked with a green circle. The bipolar outflow of IS-1, reported in \cite{duarte13,duarte14}, is also clearly visible. 
SiO emission appears to cover most of the extent of the FWHM of N48, covering both the proposed N48 outflow, the additional red-shifted lobe to the south and the blue-shifted lobe of IS-1. 

\paragraph{N50:}
HCO$^+$ emission in N50 is dominated by blue-shifted material extending southwards from DR21 Main. Even so, a clear bipolar outflow can be seen at the western edge of the FWHM of N50. Given the significant overlap of the two lobes, it is likely that the outflow extends along the line of sight. As the continuum emission in the region is strongly affected from sidelobes arising from DR21 Main, it is not possible to confirm the location of the driving source of the N50 outflow. 

\paragraph{N51:}
High-velocity HCO$^+$ emission reveals a bipolar outflow, matching the location and direction of the CO outflow \citep{Skretas22}. Additional fragmented red- and blue-shifted emission is seen at the Southeast part of N51. Given its fragmented nature, and the lack of close connection with any prominent continuum peaks in N51, this emission is not considered to be part of an outflow. Finally, an extended, filamentary, blue-shifted structure can be seen, extending from the top of N51 toward the North. This structure was previously seen in CO emission by \cite{Schneider10}. The lack of a red-shifted counterpart and of a close association with any of the reported cores, suggest that this structure is not part of a protostellar outflow.    

\paragraph{N53 \& N54:}
A prominent bipolar outflow is evident in HCO$^+$, matching the orientation of the CO outflow from \cite{Skretas22}. An additional blue-shifted structure is detected, which extends towards the South. This structure matches the one seen North of N51, and as discussed previously, does not appear to be part of an outflow. A secondary continuum peak, associated with N54, can be seen at the Southeast edge of N53. Similar to the CO maps \citep{Skretas22}, a small-scale bipolar outflow is originating from N54. In the velocity range adopted to create the map of N53, the blue-shifted emission of N54 is blended with the more widespread blue lobe coming from N53. To avoid including the N54 outflow in the calculation of outflow properties of N53, we use a higher inner velocity limit for the blue-shifted outflow, thus restricting its extent to the point that the two outflows can be separated. In turn this might lead to a slight underestimate of the outflow properties of the N53 outflow. The velocity range in the map of N54 has been adjusted to higher velocities in order to more clearly separate the two outflow lobes.  

\paragraph{N55:}
N55 is one of the sources displaying the \lq shelf'-like emission feature. The integrated intensity map reveals that this component is widespread and fragmented, and does not form part of a coherent structure. Combined with the complete lack of any blue-shifted emission, this suggests that the red-shifted emission seen in N55 is not part of any outflow, but is rather associated with widespread foreground material. Since there is no outflow emission detected, the map for N55 is shown in Fig. D.1.  

\paragraph{N56 \& N57:}
The cores N56 and N57 are located in a structure located $\sim1\arcsec$ East of the DR21 ridge. This structure is seen at a velocity of $\sim$18 km s$^{-1}$ and is likely the origin of the third velocity component detected in some of the DR21 cores discussed in Section \ref{sec:spectra}. The HCO$^+$ map of N56 reveals a clear bipolar outflow. The red lobe of the outflow is very well defined, but the blue emission appears more fragmented and noisy. The most likely reasons are the proximity of the sources at the edge of the FOV of the CASCADE coverage, but also blending of the blue-shifted outflow wing with the other velocity components (at -3 and 9 km s$^{-1}$) seen in DR21. Regardless, two possible directions can be seen for the blue-shifted lobes, hinting at the possibility of multiple outflows being driven from N56. Finally, SiO emission is detected over both sides of the blue-shifted emission, but not over the red-shifted lobe. The strongest SiO emission is associated with the blue-shifted lobe extending towards the east, and thus this direction will be adopted for the calculation of relevant outflow parameters, such as outflow extent and dynamical time. In the case of N57, there is a tentative detection of a red-lobe, which lacks any clear blue-shifted counter part. In addition, due to the even closer proximity of N57 to the edge of the field observed in CASCADE, the emission is dominated by noise, making the identification of an outflow impossible.

\section{Description of HCO$^+$ emission in remaining sources}
We provide here a short description for sources that displayed wings or other notable features in their spectra (see Sec. \ref{sec:spectra}), but were found not to drive any molecular outflows. Figure D.1 shows the corresponding integrated intensity maps. 
\label{app:sourcedescriptions}
\paragraph{DR21-22:}
DR21-22 is a prime example of sources whose spectrum displayed the \lq shelf'-like structure (Section \ref{sec:spectra}), and extending between 0 and 10 km s$^{-1}$. The integrated intensity map, covering the velocity range of the \lq shelf', reveals no coherent structures. This suggests that it is not part of an outflow, but it is rather associated with foreground material, possibly from the 9 km s$^{-1}$ component. The lack of blue-shifted emission in DR21-22 is in agreement with the absence of any outflow. 
\paragraph{N34 \& N35:}
The cores N34 and N35 are located in a dense structure West of DR21 Main, first detected by \cite{Plambeck1990}. This structure appears to be actively interacting with the outflow of DR21 Main \citep{Skretas23,Karska2025}. N34 and N35 lie in the path of the DR21 Main outflow, which is dominating their integrated intensity maps. As a result, we cannot determine the presence of any additional outflows, originating from within N34 or N35. The lack of any clear outflow activity from these cores, even though their respective spectra show impressive line wings, highlights the danger of relying solely on spectral information for the identification of molecular outflows. This is especially true in clustered regions, like Cygnus-X, where multiple cores are located in a relatively small area.         
\paragraph{N36:}
The continuum emission in the field of N36 is dominated by emission from N44, barely visible at the left bottom of the field. As a result, the weaker continuum emission associated with N36 is not clearly seen. The HCO$^+$ integrated intensity contours show fragmented red parts, corresponding to the broad but weak red-shifted component seen in the spectrum. The lack of a clearly defined lobe, and of a blue-shifted counterpart, suggests that there is no outflow associated with N36. In contrast, the outflow of N40 can be seen North of N36, directly outside the FWHM radius. Interesting to note is the presence of extended SiO emission over the core. The origin of this SiO is unclear, but is likely associated with N41, located to the East of N36.
\paragraph{N41:}
While there is some blue- and red-shifted emission near the location of N41, as seen also in the corresponding spectrum (see Fig. G.2), channel maps analysis shows that the blue-shifted emission appears significant only for velocities close to the source velocities, while the red-shifted emission is likely part of the extended red-shifted structures extending on the N - S direction. Therefore, it appears that N41 drives no significant HCO$^+$ outflow. Interestingly, significant SiO emission, at low velocities, can be seen, associated with the location of N41. This is the SiO emission that extends also over N36.
\paragraph{N43:}
In the case of N43, a bipolar outflow has been reported using CO 3-2 emission \citep{Ching18}. 
In addition, H$_2$O maser emission, often associated with outflowing material, has been reported near the continuum peak of N43. Some high-velocity HCO$^+$ components appear in the vicinity of N43. The extended blue-shifted emission, at the north of N43 matches the direction of the blue-shifted CO lobe in \cite{Ching18}, but the structure is not consistent, at higher velocities, with the origin from N43. An additional blue-shifted component, located at the Eastern edge of the FWHM of N43, does not match the CO emission, and is significantly disconnected from the continuum peak. At the same time, there is only a marginal detection of the red-shifted emission coinciding with the FWHM of N43. Therefore, it appears that there is no clear HCO$^+$ counterpart to the CO outflow in N43. Finally, significant SiO emission is detected near the continuum peak of N43. This emission appears at source velocity, and while it is relatively narrow in velocity space, it is likely to be originating in shocks related to the CO outflow.

\end{document}